\documentclass[%
 reprint,
 amsmath,amssymb,
 aps,
 prb,
]{revtex4-2}

\usepackage{graphicx}
\usepackage{dcolumn}
\usepackage{bm}
\usepackage[colorlinks = true,
            linkcolor = blue,
            urlcolor  = blue,
            citecolor = blue,
            anchorcolor = blue]{hyperref}
\usepackage{hyperref}
\usepackage{mathptmx}
\usepackage{siunitx}
\usepackage{mathptmx}
\usepackage[percent]{overpic}

\begin{document}

\preprint{APS/123-QED}

\title{Observation of intertwined charge density wave order and superconductivity in Janus monolayer}

\author{Subhajit Pramanick\textsuperscript{1,3}}
\email{subhajitbhu@kgpian.iitkgp.ac.in}
\author{Shubham Patel\textsuperscript{2}}
\email{pshubham2805@gmail.com}
\author{Sudip Chakraborty\textsuperscript{3}}
\email{sudipchakraborty@hri.res.in}
\author{A. Taraphder\textsuperscript{1}}
\email{arghya@phy.iitkgp.ac.in}

\affiliation{\textsuperscript{1}Department of Physics, Indian Institute of Technology Kharagpur, Kharagpur 721302, India}
\affiliation{\textsuperscript{2}Department of Physics, University of Bath, Claverton Down, Bath BA2 7AY, United Kingdom}
\affiliation{\textsuperscript{3}Materials Theory for Energy Scavenging Lab, Harish-Chandra Research Institute, A CI of Homi Bhabha National Institute, Chhatnag Road, Jhunsi, Prayagraj 211019, India}

\date{\today}

\begin{abstract}
Low-dimensional transition-metal dichalcogenides (TMDCs) provide an ideal platform for studying the emergence of charge density wave (CDW) and superconductivity. The discovery of emergent CDW order in 1T $\mathrm{ZrTe_2}$ monolayer raises an important question: does this instability persist when one $\mathrm{Te}$ chalcogen layer is substituted by $\mathrm{Se}$? In the present work, we investigate the CDW (2$\times$2$\times$1) and superconducting instability in 1T $\mathrm{ZrSeTe}$ Janus monolayer using first-principles calculations. The phonon spectrum exhibits a pronounced anomaly at the $\mathrm{M}$ point of the irreducible Brillouin zone, arising from enhanced electron-phonon interaction together with electronic instabilities originating from both interband and intraband scattering. The resulting lattice distortion reconstructs the electronic structure, opening a small indirect band gap, driving the system from a semi-metallic to a semiconducting state. The energy gain associated with the distortion is significantly smaller than that of $\mathrm{ZrTe_2}$ monolayer, indicating that the replacement of one $\mathrm{Te}$  chalcogen layer with $\mathrm{Se}$ weakens the CDW instability. We have further investigated the effects of electronic correlation and biaxial strain, both acts as effective tuning parameters for the instabilities concerened. In the high temperature undistorted phase, $\mathrm{ZrSeTe}$ exhibits phonon mediated two-gap superconductivity. It originates primarily from the robust coupling between the soft phonon mode at $\mathrm{M}$ point and the electronic bands predominantly derived from $\mathrm{Zr}$ \textit{d} and $\mathrm{Te}$ \textit{p} orbitals crossing the Fermi level. Spin-orbit coupling (SOC) further modifies the electronic states and reduces the superconducting transition temperature.

\end{abstract}


\maketitle


\section{\label{sec:1}Introduction}

In many materials, the interplay among different electronic instabilities gives rise to fascinating emergent phenomena. Among them, CDW and superconductivity are two typical Fermi surface instabilities that have attracted significant interest due to their competing nature as well as potential coexistence. From a materials perspective, Cuprates \cite{chang2012direct,da2014ubiquitous}, TMDCs \cite{wilson1974charge,wilson1975charge,halperin1968possible,halperin1968excitonic,lian2022intrinsic,zheng2019electron,zhang2023emergent,shao2016manipulating,wei2017manipulating,ren2022semiconductor,patel2024electron}, some classes of kagome metals \cite{tan2021charge,lin2022multidome,cao2023competing}, etc., are promising candidates for exhibiting emergent phenomena like CDW, superconductivity, etc. Historically, three major mechanisms are suggested for the formation of CDW \cite{peierls1955quantum,kohn1959image,zhu2015classification,kaboudvand2022fermi,pramanick2025pressureinducedevolutionanisotropic}. The first mechanism follows the Peierls scenario, where the instability originates from the Fermi surface nesting (FSN). The second mechanism is driven by the momentum dependent electron-phonon coupling (EPC). The last mechanism is the electron-electron correlation in correlated system such as cuprates, nickelates where FSN or EPC are not sufficient to explain the putative charge ordering\cite{peierls1955quantum,kohn1959image,zhu2015classification,kaboudvand2022fermi,pramanick2025pressureinducedevolutionanisotropic}. Apart from this, alternative mechanisms \cite{jerome1967excitonic,taraphder2011preformed,koley2015unusual,koley2020charge,cercellier2007evidence,monney2011exciton,wegner2020evidence,wang2022origin,yin2020theoretical}, based primarily on exitonic-liquid scenarios, have been proposed lately to address the broader question of the formation of CDW without pronounced FS signatures, from an anomalous normal state (bad metal) in several TMDs. The nature of CDWs in two dimensional TMDCs is probably more complex than that of many quasi-one-dimensional systems, such as $\mathrm{ZrTe_3}$, $\mathrm{NbSe_3}$, and certain organic compounds, where it is derived primarily from FSN and follows the Peierls picture \cite{ren2022semiconductor,gabovich2002charge,hoesch2009giant,monceau2012electronic}. The conflict between CDW and superconductivity in various monolayer and bilayer TMDCs has been addressed in several theoretical and experimental studies conducted over the past few years \cite{lian2022intrinsic,zheng2019electron,wei2017manipulating,patel2024electron}. $\mathrm{TaS_2}$ is such a notable TMD that can display various CDW phases in its 1T structure, and a coexistence of superconductivity and CDW phases in its 2H structure \cite{lian2022intrinsic}. Experiments on 1T-$\mathrm{TaS_2}$ have shown a series of incommensurate to commensurate CDW transitions, with the formation of a “Star-of-David" cluster and a corresponding Mott insulating state characterizing the low-temperature phase \cite{wang2020band,cho2017correlated,ma2016metallic}. On the other hand, 2H $\mathrm{TaS_2}$ exhibits a 3$\times$3 CDW transition followed by superconductivity, where the instability is primarily driven by momentum-dependent electron-phonon interaction rather than simple Fermi surface nesting \cite{joshi2019short,wijayaratne2017spectroscopic}. First principles calculations on 1H $\mathrm{TaS_2}$ monolayer show that electron doping gradually shifts the phonon instability from $\frac{2}{3}\mathrm{\Gamma M}$ towards $\mathrm{M}$ point, driving a transition from $\mathrm{3\times3}$ to $\mathrm{2\times2}$ CDW instability, while simultaneously enhancing superconductivity and inducing a crossover from single gap to two gap behavior \cite{lian2022intrinsic}. In monolayer 2H $\mathrm{NbSe_2}$, superconductivity and CDW order also coexist, believed by some to be closely connected through EPC \cite{zheng2019electron}. It is argued that in this system, intra-pocket scattering favors superconductivity while inter-pocket scattering promotes CDW formation and partial gapping of the Fermi surface \cite{zheng2019electron}. While 1T $\mathrm{VTe_2}$ exhibits multiple competing CDW phases, including 4$\times$4 and $\mathrm{2\sqrt{3}\times2\sqrt{3}}$, characterized by anisotropic gap formation and phonon softening, with the instability arising from the combined effects of FSN and momentum-dependent EPC \cite{wang2019evidence,wang2025unraveling}, transport in 1T $\mathrm{VSe_2}$ CDW state shows an anomaly at 20K and an isosbestic point in the Seebeck effect with magnetic field \cite{Sonika_2025}.

\par In addition to group V TMDCs, group IV-based TMDCs also host prominent CDW instabilities \cite{zhang2023emergent,wei2017manipulating,ren2022semiconductor,cercellier2007evidence,monney2011exciton,wegner2020evidence}. 1T $\mathrm{TiTe_2}$ exhibits a remarkable dimensional crossover, where a 2$\times$2 CDW phase emerges in the monolayer limit, accompanied by a pseudogap near the Fermi level, while the bulk remains free of CDW order  \cite{chen2017emergence,sky2020theory}. 1T $\mathrm{TiSe_2}$ has been widely studied due to an unconventional CDW and rich electronic properties. The origin of CDW in semi-metallic 1T $\mathrm{TiSe_2}$, with low carrier density, remains a subject of ongoing debate with several competing scenarios proposed in the literature, including exciton driven ordering \cite{koley2014preformed,cercellier2007evidence,monney2011exciton}, Jahn-Teller types lattice instabilities \cite{wegner2020evidence,rossnagel2002charge}, electron-phonon interaction \cite{weber2011electron,hellgren2017critical} and correlation induced effects \cite{novko2022electron}. $\mathrm{T}$-phase of $\mathrm{ZrTe_2}$ is another promising member of the TMDC family that has recently attracted considerable attention. In monolayer limit, it exhibits a 2$\times$2 CDW instability, as verified through scanning tunneling microscopy (STM) and scanning tunneling spectroscopy (STS) measurements \cite{yang2022coexistence}, angle-resolved photoemission spectroscopy (ARPES) measurements and fast Fourier transform (FFT) analysis of the monolayer $\mathrm{ZrTe_2}$ grown by molecular beam epitaxy (MBE) technique \cite{ren2022semiconductor,song2023signatures}. First-principles calculations further confirm it through the observation of phonon softening at the $\mathrm{M}$ point and a pronounced energy-gap opening in the outer valence band along the $\mathrm{\Gamma-M}$ path \cite{zhang2023emergent}. Recent experimental studies on ultrathin 1T $\mathrm{ZrX_2}$ ($\mathrm{X= Se, Te}$) have revealed a 2$\times$2 CDW in the two-dimensional limit, emerging after a semiconductor to metal transition induced by charge transfer from the substrate \cite{ren2022semiconductor}. These findings highlight the importance of tuning the Fermi level to induce metallicity, which enables CDW formation driven by electronic instability. In this context, the Janus $\mathrm{ZrSeTe}$ monolayer is important, as its intrinsic asymmetry can modify the electronic structure and change the vibrational properties \cite{dey2020structural}, providing a platform to investigate and control the CDW instability.

\par In the present study, we perform first-principles calculations to investigate the CDW instability, its microscopic origin, and the associated electronic reconstruction in the Janus TMDC 1T $\mathrm{ZrSeTe}$. The lattice dynamical calculations reveal the presence of a triple \textit{q} phonon instability at the $\mathrm{M}$ point, indicating a strong tendency toward CDW formation. A detailed analysis of the electronic instability and the electron-phonon interaction demonstrates their crucial roles in driving the instability. Using the frozen-phonon calculations, we determine the distorted ground state and find that the energy gain of the corresponding CDW distortion becomes significantly smaller than that of the non-Janus counterpart, the $\mathrm{ZrTe_2}$ monolayer. We also explore the influence of electronic correlation and biaxial strain on the CDW instability. In addition, we investigate the superconducting properties of high temperature undistorted phase, where electron-phonon interaction is expected to play a significant role. Within the Migdal-Eliashberg formalism, our results reveal a phonon mediated anisotropic two-gap superconductivity in this system.

\section{\label{sec:2}Methodology}

Density functional theory are performed employing the Perdew-Burke-Ernzerhof (PBE) exchange-correlation functional \cite{perdew1996generalized} within the generalized gradient approximation to conduct electronic structure calculations. The projector augmented-wave (PAW) method \cite{blochl1994projector,kresse1999ultrasoft} implemented in the Vienna Ab initio Simulation Package (VASP) \cite{kresse1993ab,kresse1996efficiency} has been used for this purpose. Complete ionic relaxation is performed via the conjugate gradient algorithm until the force on each atom is below -0.001 eV/\text{\AA}. An energy tolerance limit of $\mathrm{10^{-8}}$ eV and an energy cutoff of 300 eV are used for this. Density Functional Perturbation Theory (DFPT) implemented in the Quantum Espresso package \cite{giannozzi2009quantum,giannozzi2017advanced,giannozzi2020quantum} has been utilized to study electronic structure and phonon dynamics using scalar relativistic ultrasoft pseudopotentials from Garrity-Bennett-Rabe-Vanderbilt (GBRV) library \cite{garrity2014pseudopotentials} and full relativistic ultrasoft pseudopotentials from PSlibrary \cite{giannozzi2009quantum,giannozzi2017advanced,giannozzi2020quantum} of PBEsol functional. For calculations performed without and with SOC, kinetic energy cutoffs of 30 Ry and 70 Ry, and charge-density cutoffs of 240 Ry and 550 Ry, respectively, have been used. Electronic occupations are treated using cold smearing with a width of 0.005 Ry unless otherwise noted. However, the superconductivity of the high temperature, undistorted phase has been calculated using a smearing width of 0.01 Ry. The phonon dispersion curves of the high symmetry undistorted phase have been constructed via Fourier interpolation of the dynamical matrices computed on a 24 $\times$ 24 $\times$ 1, 12 $\times$ 12 $\times$ 1 \textit{k}-point meshes and 12 $\times$ 12 $\times$ 1, 6 $\times$ 6 $\times$ 1 \textit{q}-point meshes without and with SOC considerations respectively. On the other hand, phonon dispersion of 2$\times$2 CDW distorted phase has been calculated using a 10 $\times$ 10 $\times$ 1 \textit{k}-point mesh and 5 $\times$ 5 $\times$ 1 \textit{q}-point mesh. The Allen-Dynes modified McMillan formula \cite{mcmillan1968transition,allen1975transition} and the Migdal-Eliashberg formalism \cite{migdal1958interaction,eliashberg1960interactions} implemented in Electron Phonon Wannier (EPW) \cite{ponce2016epw,margine2013anisotropic,giustino2007electron} code are employed to calculate EPC and superconducting properties. The superconducting critical temperature in the isotropic regime has been calculated using the following Allen-Dynes modified McMillan formula \cite{mcmillan1968transition,allen1975transition}:
\begin{eqnarray}
\mathrm{T_c = \frac{{\omega}_{log}}{1.20} \hspace{0.1cm} exp\left[\frac{-1.04 \hspace{0.1cm} (1 + \lambda)}{\lambda - {\mu}_c^{\ast} (1+0.62\lambda)}\right]} 
\label{eq:1}
\end{eqnarray}
Where $\mathrm{\omega_{log}}$, $\mathrm{\lambda}$, and $\mathrm{\mu_c^*}$ are the logarithmic average phonon frequency, isotropic EPC strength, and effective Coulomb potential, respectively. Isotropic cumulative EPC strength is
\cite{ponce2016epw,margine2013anisotropic}:
\begin{eqnarray}
\mathrm{\lambda = 2 \int_{0}^{\omega} \frac{\alpha^2 F(\omega)}{\omega} d\omega}
\label{eq:2}
\end{eqnarray}
Isotropic Eliashberg electron-phonon spectral function $\mathrm{\alpha^2 F(\omega)}$ can be calculated by:
\begin{eqnarray}
\mathrm{\alpha^2 F(\omega) = \frac{1}{2\pi N_F} \sum_{\textbf{q},\nu} \frac{\gamma_{\textbf{q}\nu}}{\omega_{\textbf{q}\nu}} \delta (\omega - \omega_{\textbf{q}\nu})}   
\label{eq:3}
\end{eqnarray}
where, $\mathrm{\gamma_{q\nu}}$, $\mathrm{\omega_{q\nu}}$ and $\mathrm{N_F}$ represent the linewidth of the phonon associated with branch index $\mathrm{\nu}$ and momentum $\mathbf{q}$, the phonon frequency of the corresponding mode and momenta, and the electron density of states at the Fermi level respectively. On the other hand, the superconducting critical temperature in the anisotropic regime has been calculated from the self-consistent solution of the following Migdal-Eliashberg equations \cite{migdal1958interaction,eliashberg1960interactions}:
\begin{eqnarray}
\mathrm{Z(\textbf{k}_1, i\omega_{n_{1}}) =} && \mathrm{1 + \frac{\pi T}{N_F \omega_{n_{1}}} \sum_{\textbf{k}_2, n_2} \frac{\omega_{n_{2}}}{\sqrt{\omega_{n_{2}}^2 + \Delta^2(\textbf{k}_2, i\omega_{n_{2}})}}}  \nonumber\\ 
&& \mathrm{\times \lambda(\textbf{k}_1, \textbf{k}_2, n_1 - n_2) \delta(\epsilon_{\textbf{k}_{2}})}  
\label{eq:4}
\end{eqnarray}
\begin{eqnarray}
\mathrm{Z(\textbf{k}_1,} &&\mathrm{i\omega_{n_{1}}) \Delta (\textbf{k}_1, i\omega_{n_{1}}) = \frac{\pi T}{N_F} \sum_{\textbf{k}_2, n_2} \frac{\Delta (\textbf{k}_2, i\omega_{n_{2}})}{\sqrt{\omega_{n_{2}}^2 + \Delta^2(\textbf{k}_2, i\omega_{n_{2}})}}}\nonumber\\ 
&& \mathrm{\times [\lambda (\textbf{k}_1, \textbf{k}_2, n_1 - n_2) - N_F V(\textbf{k}_1 - \textbf{k}_2)] \delta (\epsilon_{\textbf{k}_{2}})}
\label{eq:5}
\end{eqnarray}

In Eqs. \hyperref[eq:4]{4} and \hyperref[eq:5]{5}, $\mathrm{Z}$, $\mathrm{\Delta}$, V$(\mathbf{k_1 - k_2})$, $\mathrm{i\omega_n}$ (= $\mathrm{i(2n+1)\pi T}$ for integer n) are
renormalization function, superconducting gap, screened Coulomb interaction, and fermionic Matsubara frequencies, respectively. The upper limit of the frequency integration in the Eliashberg equations is set to 0.15 eV, which is five times the maximum phonon frequency in the phonon dispersion. Wannier basis to Bloch basis interpolation for EPC calculation using EPW \cite{ponce2016epw,margine2013anisotropic,giustino2007electron} are carried out on a fine 600 $\times$ 600 $\times$ 1 \textit{k}-point grid and a 120 $\times$ 120 $\times$ 1 \textit{q}-point grid. The details of Wannier projections are provided in the supplemental material \cite{supplementary}. For the calculation of Fermi surface nesting and EPC matrix element, a 600$\times$600$\times$1 and 60$\times$60$\times$1 fine \textit{k}-grids, respectively, have been adopted. The smearing width for the energy-conserving delta function and the width of the Fermi window are set to 0.03 eV and 0.12 eV, respectively. The effective Coulomb potential is taken as 0.1 eV. The crystal structures and Fermi surfaces have been visualized with the VESTA \cite{momma2011vesta} and FermiSurfer \cite{kawamura2019fermisurfer} tools. The interface between PyProcar \cite{Herath2020107080, Lang2024109063} and VASP \cite{kresse1993ab,kresse1996efficiency} has been utilized to plot unfolded band structure for the 2$\times$2 distorted phase.

\section{\label{sec:3}Results and Discussions}

\subsection{\label{sec:3.1}Study of undistorted phase}

\begin{figure}[h!]
\includegraphics[width=1.0\linewidth]{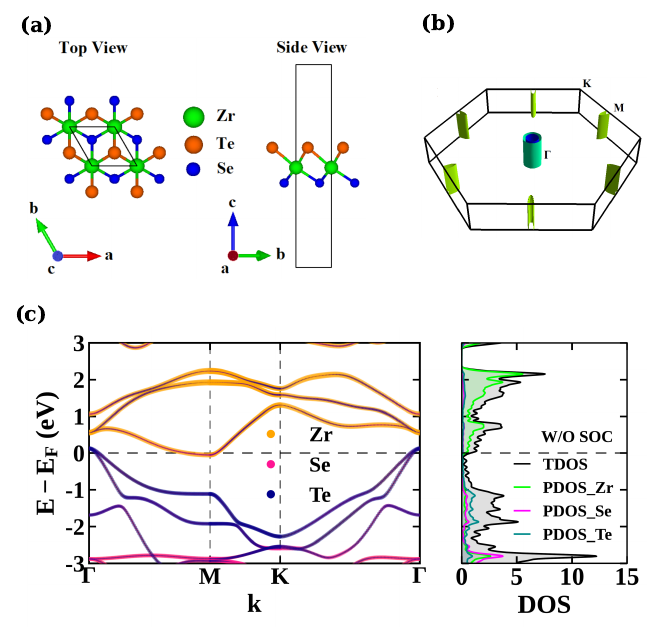}

\caption{\label{fig:1} (a) Crystal structure with its top and side view, (b) Fermi surface, (c) electronic band structure and density of states of 1 T $\mathrm{ZrSeTe}$ monolayer at ambient conditions. SOC has not been considered here.}
\end{figure}

At ambient conditions, a monolayer of the T phase of $\mathrm{ZrSeTe}$, as shown in Fig. \hyperref[fig:1]{1(a)}, belongs to a two-dimensional hexagonal (hp) type Bravais lattice having p3m1 layer group, where Zr is sandwiched between Se and Te layers. Due to having two different chalcogen layers, the 1T $\mathrm{ZrSeTe}$ monolayer exhibits a non-centrosymmetric crystal structure, breaking inversion symmetry. The calculated formation energy \cite{peterson2021materials} of 1T $\mathrm{ZrSeTe}$ monolayer is -1.12 eV/atom, indicating that the structure is energetically favorable. The negative formation energy suggests that the monolayer can be stable with respect to its constituent elements. Fermi surface and electronic dispersion of the monolayer $\mathrm{ZrSeTe}$ calculated using PBE functional \cite{perdew1996generalized} in the absence of SOC have been displayed in Figs. \hyperref[fig:1]{1(b-c)}. It exhibits semimetallic behavior, with three bands intersecting the Fermi level. Two valence bands, primarily derived from $\mathrm{Te}$ $\textit{p}$ orbitals and one conduction band, dominated by $\mathrm{Zr}$ $\textit{d}$ orbitals, intersect the Fermi level near $\mathrm{\Gamma}$ point and at the $\mathrm{M}$ point respectively. Two hole pockets having $\mathrm{Te}$ $\textit{p}$ orbitals characteristics at $\mathrm{\Gamma}$ point and an elongated elliptical electron pocket

having $\mathrm{Zr}$ $\textit{d}$ orbitals characteristics at $\mathrm{M}$ point are clearly seen in the corresponding Fermi surface plot of Fig. \hyperref[fig:1]{1(b)}. In the presence of SOC, using the PBE functional \cite{perdew1996generalized}, the calculated electronic band structure and corresponding Fermi surface have been displayed in Fig. S1(b-c) of the supplemental material \cite{supplementary}. The Fermi energy decreases slightly as a result of the SOC. $\mathrm{Zr}$ $\textit{d}$ orbitals dominated conduction band which crosses the Fermi level at $\mathrm{M}$ point, now moves towards the conduction bands. The degeneracy of two valence bands at the $\mathrm{\Gamma}$ point is lifted by the strong atomic SOC of $\mathrm{Te}$ atoms. However, at the $\mathrm{\Gamma}$ point, both bands remain doubly degenerate due to Kramers degeneracy, which is protected by time reversal symmetry (TRS). These lead to the disappearance of one hole pocket at $\mathrm{\Gamma}$ point and the electron pocket at $\mathrm{M}$ point from the Fermi surface. Due to the absence of inversion symmetry, SOC lifts the spin degeneracy of several conduction and valence bands away from the time-reversal invariant momenta (TRIM). However, at the TRIM points ($\mathrm{\Gamma}$ and $\mathrm{M}$), Kramers degeneracy enforced by TRS preserves the band degeneracy for up and down spins. It is noteworthy that the observed spin splitting occurs outside the Fermi window considered for EPC calculations, and therefore does not significantly influence EPC. The band overlap in these semimetallic systems can be quantified by defining a negative band gap as the energy difference between the minima of the conduction band at the $\mathrm{M}$ point and the maxima of the valence band at the $\mathrm{\Gamma}$ point \cite{zhang2023emergent}. Our calculated negative band gap for the Janus $\mathrm{ZrSeTe}$ monolayer is smaller than that of the 1T $\mathrm{ZrTe_2}$ monolayer \cite{zhang2023emergent}. It can be attributed to the difference in $\mathrm{Se}$ and $\mathrm{Te}$ orbital energies and the structural asymmetry of the Janus layer that shifts the band edges and reduces the band overlap near the Fermi level \cite{zhao2025preparation}.

\begin{figure*}[t]
\includegraphics[width=0.75\linewidth]{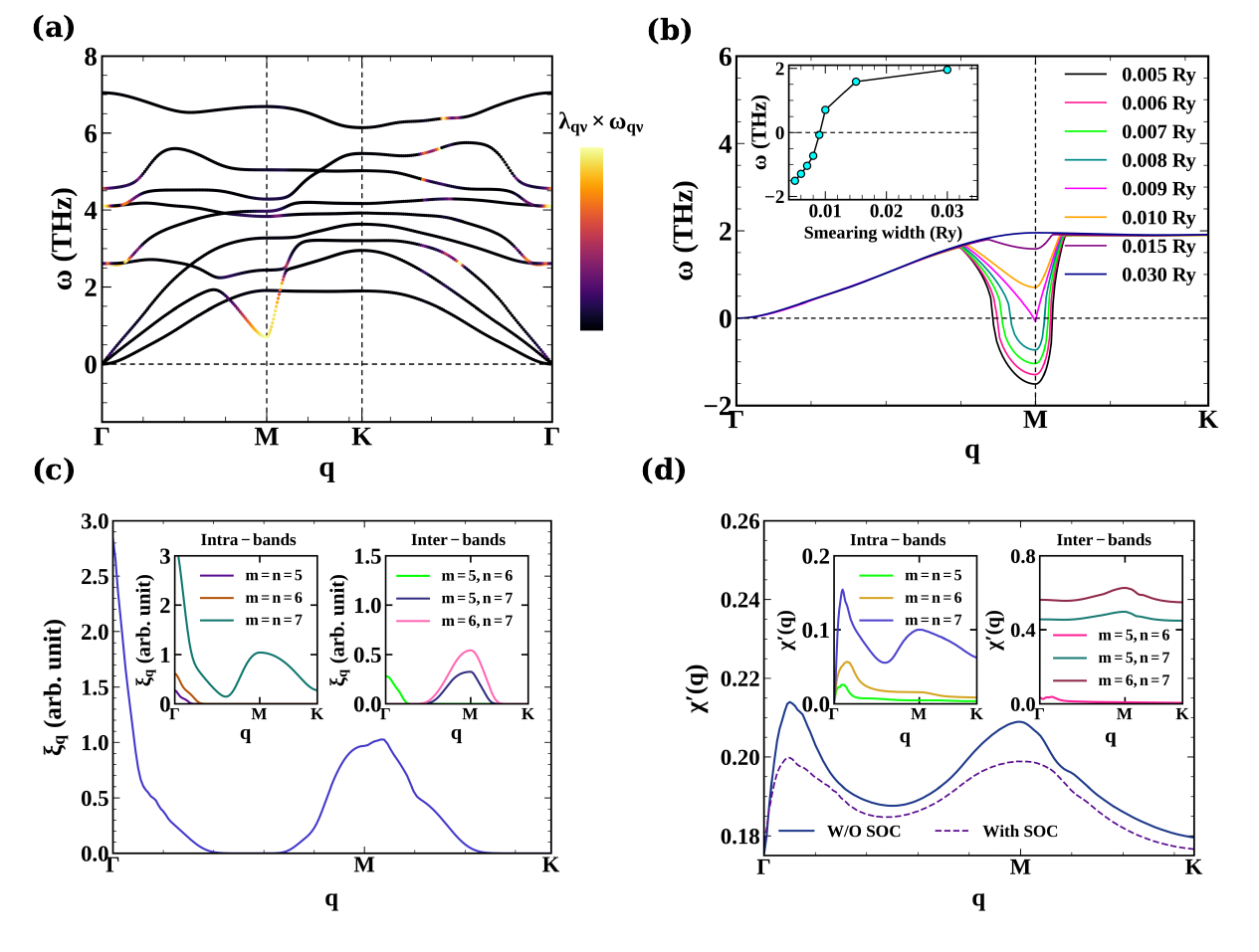}

\caption{\label{fig:2} (a) Phonon dispersion and mode resolved $\mathrm{\lambda_{q\nu}\times\omega_{q\nu}}$ along high symmetry $\textit{q}$ points of $\mathrm{ZrSeTe}$ monolayer in absence of SOC; (b) Main: Phonon softening of an acoustic mode under different values of smearing parameter, i.e., temperature, Inset: Phonon frequency of the soft acoustic mode at $\mathrm{M}$ point vs different values of smearing parameter; (c) Imaginary part and (d) Real part of the bare static Lindhard charge susceptibility along $\mathrm{\Gamma-M-K}$ path for, Main: total contribution of bands that cross the Fermi level and Inset: intra- and inter-bands contribution.}
\end{figure*}

\subsection{\label{sec:3.2}Signature of CDW instability and its mechanism}

CDW state emerges from a spontaneous symmetry breaking transition characterized by periodic modulations of the electronic charge density and accompanying lattice distortions below a critical temperature $\mathrm{T_{CDW}}$ \cite{wang2023decisive}. In a phonon-driven CDW system, the instability is signaled by the softening of a phonon mode at a characteristic wave vector \cite{lian2022intrinsic,zheng2019electron,zhang2023emergent,wang2023decisive}. Under the random phase approximation, the frequency ($\mathrm{\omega_{\mathbf{q},\nu}}$) of softened phonon corresponding to wave vector $\mathbf{q}$ and mode $\mathrm{\nu}$ can be written as \cite{wang2023decisive}:
\begin{eqnarray}
\mathrm{\omega_{\mathbf{q},\nu}^2 = \Omega_{\mathbf{q},\nu}^2 - 2\Omega_{\mathbf{q},\nu} \chi_{\mathbf{q},\nu}}
\label{eq:6}
\end{eqnarray}
Where $\mathrm{\Omega_{\mathbf{q},\nu}}$ and $\mathrm{\chi_{\mathbf{q},\nu}}$ represent bare phonon frequency and the generalized static electronic susceptibility. Figure \hyperref[fig:2]{2(a)} presents the phonon dispersion of the $\mathrm{ZrSeTe}$ monolayer calculated without including SOC, using a smearing width of 0.01 Ry. The results indicate that one of the acoustic branches exhibits a pronounced tendency toward softening at the $\mathrm{M}$ point, suggesting a 2$\times$2 CDW order. The evolution of this soft phonon mode with respect to the electronic temperature, represented here by the smearing width, is shown in the main panel and the inset of Fig. \hyperref[fig:2]{2(b)}. As the smearing width approaches 0.009 Ry, a distinct discontinuity appears in the phonon dispersion, indicating the presence of a Kohn anomaly \cite{yin2020theoretical}. Soft mode in phonon dispersion of $\mathrm{ZrSeTe}$ monolayer calculated including SOC has been displayed in Fig. S1(d) of the supplemental material \cite{supplementary}. For a multi band case, $\mathrm{\chi_{\mathbf{q},\nu}}$ in Eq. \hyperref[eq:6]{6} can be formulated as \cite{wang2023decisive}:
\begin{eqnarray}
\mathrm{\chi_{\mathbf{q},\nu} = \sum_{m,n,k} |g_{m,n,\nu}(\mathbf{k},\mathbf{q})|^2\frac{f(\epsilon_{n\textbf{k}})-f(\epsilon_{m\textbf{k}+\textbf{q}})}{\epsilon_{m\textbf{k}+\textbf{q}}-\epsilon_{n\textbf{k}}}}
\label{eq:7}
\end{eqnarray}
Here, $\mathrm{g_{m,n,\nu}(\mathbf{k},\mathbf{q})}$ is the EPC matrix element corresponding to the coupling between electronic states $\mathbf{k}$ of band index $\mathrm{n}$ and electronic states $\mathbf{k}+\mathbf{q}$ of band index $\mathrm{m}$ with a phonon having momentum $\mathbf{q}$, mode $\mathrm{\nu}$. $\mathrm{f}$ is the Fermi-Dirac function and $\mathrm{\epsilon_{n\textbf{k}}}$, $\mathrm{\epsilon_{m\textbf{k}+\textbf{q}}}$ are the Kohn-Sham eigen values corresponding to band index $\mathrm{n}$, $\mathrm{m}$ and momenta $\mathbf{k}$, $\mathbf{k}$+$\mathbf{q}$ respectively. In constant fraction approximation, $\mathrm{\chi_{\mathbf{q},\nu}}$ in Eq. \hyperref[eq:7]{7} turns into the $\mathbf{q}$-EPC $\mathrm{\bar{g}_{\mathbf{q},\nu}}$ \cite{wang2023decisive}:
\begin{eqnarray}
\mathrm{\bar{g}_{\mathbf{q},\nu} = \sum_{m,n,k}|g_{m,n,\nu}(\mathbf{k},\mathbf{q})|^2}
\label{eq:8}
\end{eqnarray}
which captures the EPC effect. $\mathrm{g_{m,n,\nu}(\mathbf{k},\mathbf{q})}$ depends on the self-consistent potential corresponding to a phonon having wave vector $\mathbf{q}$, mode $\mathrm{\nu}$, and frequency $\mathrm{\omega_{\mathbf{q},\nu}}$ through the following Eq. \hyperref[eq:9]{9} \cite{ponce2016epw}:
\begin{eqnarray}
\mathrm{g_{m,n,\nu}(\mathbf{k},\mathbf{q}) = \sqrt{\frac{\hbar}{2M_i\omega_{\mathbf{q},\nu}}}\langle\psi_{m,\mathbf{k}+\mathbf{q}}|\partial_{\mathbf{q}\nu}V|\psi_{n,\mathbf{k}}\rangle}
\label{eq:9}
\end{eqnarray}
where $\mathrm{M_i}$ represents the mass of the ion, $\mathrm{\psi_{m\mathbf{k}+\mathbf{q}}}$ and $\mathrm{\psi_{n,\mathbf{k}}}$ are electronic wave functions of the corresponding band and wave vector. The EPC impact may also be observed through the projection of the mode-resolved EPC strength ($\mathrm{\lambda_{\mathbf{q}\nu}}$), which is contingent upon the EPC matrix element. The color plot of mode resolved $\mathrm{\lambda_{\mathbf{q}\nu}\times\omega_{\mathbf{q}\nu}}$, on the phonon bands has been shown in Fig. \hyperref[fig:2]{2(a)}. The presence of larger value of $\mathrm{\lambda_{\mathbf{q}\nu}\times\omega_{\mathbf{q}\nu}}$ in that soft mode around $\mathrm{M}$ point distinctly reflects the EPC effect at $\mathrm{\mathbf{q} = M \, (\frac{1}{2},0,0)}$ point. On the other hand, in constant matrix element approximation, $\mathrm{\chi_{\mathbf{q},\nu}}$ in Eq. \hyperref[eq:7]{7} turns into the real part of static Lindhard susceptibility $\mathrm{\chi_\mathbf{q}^\prime}$ \cite{wang2023decisive,johannes2008fermi,johannes2006fermi}:
\begin{eqnarray}
\mathrm{\chi_\mathbf{q}^\prime = \sum_{m,n,k} \frac{f(\epsilon_{n\textbf{k}})-f(\epsilon_{m\textbf{k}+\textbf{q}})}{\epsilon_{m\textbf{k}+\textbf{q}}-\epsilon_{n\textbf{k}}}}
\label{eq:10}
\end{eqnarray}
which reflects electronic instability associated with Fermi surface nesting. The evidence of Fermi surface nesting can be explicitly found from the presence of pronounced peaks in the imaginary part of the bare, static Lindhard charge susceptibility $\mathrm{\xi_\mathbf{q}}$ \cite{johannes2008fermi,johannes2006fermi}:
\begin{eqnarray}
\mathrm{\xi_\textbf{q} = \sum_{m,n} \int_{BZ} \frac{d\textbf{k}}{\Omega_{BZ}} \delta(\epsilon_{n\textbf{k}} -\epsilon_F)\delta(\epsilon_{m\textbf{k}+\textbf{q}} - \epsilon_F)}
\label{eq:11}
\end{eqnarray}
Here, $\mathrm{\epsilon_F}$ is the Fermi energy. Calculated $\mathrm{\chi_\textbf{q}^\prime}$ and $\mathbf{\xi_\textbf{q}}$ along the $\mathrm{\Gamma-M-K}$ path of the irreducible Brillouin zone (IBZ) have been displayed in Figs. \hyperref[fig:2]{2(c-d)}. Total $\mathrm{\xi_\mathbf{q}}$, its interband and intraband contributions in the absence of SOC have been shown in the main part and inset of Fig. \hyperref[fig:2]{2(c)}. Valence bands that cross the Fermi level near $\mathrm{\Gamma}$ point and conduction band that crosses the Fermi level at $\mathrm{M}$ point are denoted here by band indices $\mathrm{m,n = 5,6,7}$ respectively. For intraband contributions, $\mathrm{\xi_\mathbf{q}}$ exhibits pronounced peaks (self-nesting peak) at the $\mathrm{\Gamma}$ point ($\mathrm{q = 0}$), which is directly associated with the density of states at the Fermi level for each band \cite{pramanick2025pressureinducedevolutionanisotropic}. The local maxima in $\mathrm{\xi_\mathbf{q}}$ for the band with index 7 at the $\mathrm{M}$ point clearly indicates potential finite nesting between distinct regions of the same Fermi surface with the vector $\mathrm{\mathbf{q} = M(\frac{1}{2},0,0)}$. The interband contribution reveals a self-nesting peak at the $\mathrm{\Gamma}$ point for bands indexed 5 and 6, indicating their degeneracy at that point, as already captured in Fig. \hyperref[fig:1]{1(c)}. The interband contribution clearly indicates that the Fermi surface associated with band index 7 can be effectively nested with the Fermi surfaces of bands indexed 5 and 6 using the $\mathrm{\mathbf{q} = M(\frac{1}{2},0,0)}$ wavevector. However, the appearance of the peak in $\mathrm{\xi_\mathbf{q}}$ at a certain nonzero momentum does not always mean the CDW softening of phonon modes around that momentum will happen. It is the divergence or the peak (for real system) of $\mathrm{\chi_\textbf{q}^\prime}$ that matters, reflected in Eq. \hyperref[eq:6]{6}. The computed $\mathrm{\chi_\textbf{q}^\prime}$ along $\mathrm{\Gamma-M-K}$ path, displayed in Fig. \hyperref[fig:2]{2(d)} exhibits an identical peak at $\mathrm{M}$ point. It arises from both intraband scattering within band 7 and interband scattering between bands 5 and 7, and bands 6 and 7. In the presence of SOC, bands 6 and 7 no longer cross the Fermi level, as displayed in Fig. S1(b) of our supplemental material \cite{supplementary}. Consequently, the divergence of $\mathrm{\xi_\mathbf{q}}$ at the $\mathrm{M}$ point disappears since $\mathrm{\xi_\mathbf{q}}$ is governed by a double-delta function that restricts contributions to electronic states at the Fermi surface. On the other hand, even with SOC, $\mathrm{\chi_\textbf{q}^\prime}$ still shows a local maximum at the $\mathrm{M}$ point, as it also includes electronic states near the Fermi level. Therefore, the robust EPC along with an electronic instability related to intraband and interband scattering at the $\mathrm{M}$ point serves as one of the driving factors for the predicted CDW instability in $\mathrm{ZrSeTe}$ monolayer. Recent studies on monolayer 1T $\mathrm{ZrTe_2}$ have reported signatures of excitonic correlations beyond the 2$\times$2 CDW phase \cite{song2023signatures}. Therefore, it remains an open question whether excitonic effects also play a role in the instability of $\mathrm{ZrSeTe}$ monolayer.

\begin{figure*}[t]
\includegraphics[width=1\linewidth]{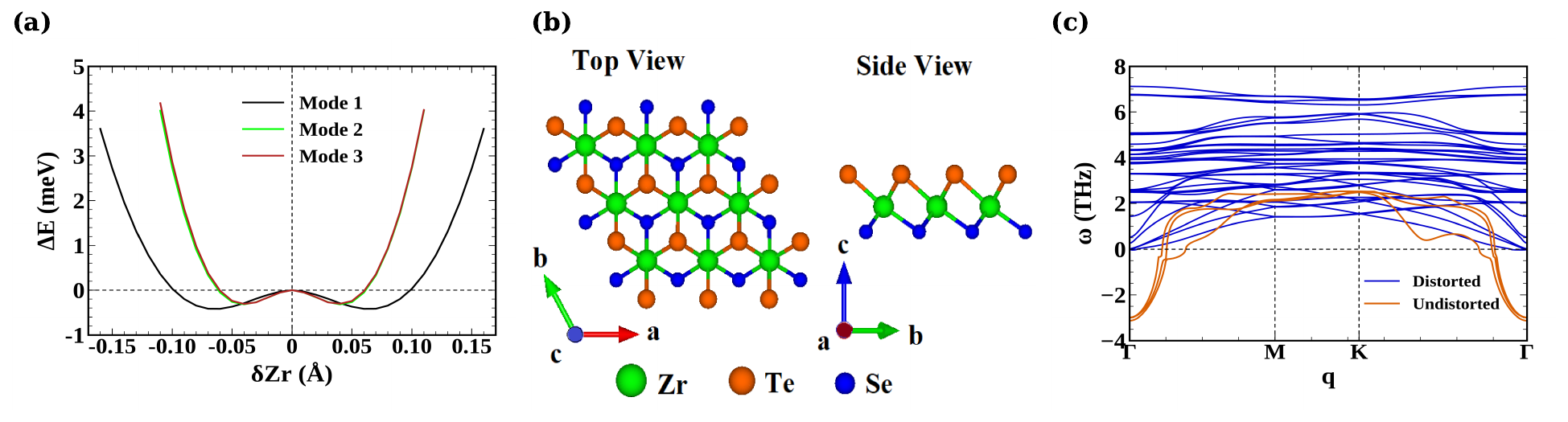}
\caption{\label{fig:3} (a) CDW double well potential formation for three unstable phonon modes of a 2$\times$2$\times$1 supercell of $\mathrm{ZrSeTe}$;  (b) Crystal structure of distorted 2$\times$2$\times$1 supercell of $\mathrm{ZrSeTe}$ with its top and side view; (c) Phonon dispersion of undistorted (orange color) and distorted (blue color) 2$\times$2$\times$1 supercell of $\mathrm{ZrSeTe}$ calculated with a smearing width 0.005 Ry. SOC has not been considered for these calculations (Figs. \hyperref[fig:3]{3(a-c)}).}
\end{figure*}

\begin{table*}[t]
\caption{\label{tab:1}%
Comparison between structures of the undistorted 2$\times$2$\times$1 $\mathrm{ZrSeTe}$ and the distorted 2$\times$2$\times$1 $\mathrm{ZrSeTe}$
}
\begin{ruledtabular}
\renewcommand{\arraystretch}{1.25}     
\setlength{\tabcolsep}{10pt} 
\begin{tabular}{ccccccccc}
2$\times$2 $\mathrm{ZrSeTe}$ & a (\AA) & b (\AA) & c (\AA) & $\mathrm{\alpha}$ & $\mathrm{\beta}$ & $\mathrm{\gamma}$ & $\mathrm{Zr-Se}$ (\AA) & $\mathrm{Zr-Te}$ (\AA)\\
\hline
Undistorted  & 7.7885 & 7.7885 & 39.3562 & \ang{90} & \ang{90} & \ang{120} &  2.6919 & 2.9343 \\
Distorted &  7.7867 &  7.7868 &  39.3706 & \ang{89.9946} & \ang{90.0054} & \ang{119.9909} & 2.6921 & 2.9345 \\
\end{tabular}
\end{ruledtabular}
\end{table*}

\subsection{\label{sec:3.3}Study of distorted phase}

In order to investigate the low-temperature CDW distorted phase, we have utilized a frozen-phonon-like approach \cite{zhang2023emergent}. Motivated by the phonon softening observed at the $\mathrm{M}$$(\frac{1}{2},0,0)$ point, which signals a tendency toward a 2$\times$2$\times$1 superstructure, we investigate the corresponding distortion in $\mathrm{ZrSeTe}$ monolayer in light of previous theoretical \cite{zhang2023emergent} and experimental \cite{yang2022coexistence,ren2022semiconductor,song2023signatures} studies. SOC has been disregarded to determine the CDW distorted phase as it has no significant impact on the order of CDW at ambient conditions. Three unstable modes have been identified at the $\mathrm{\Gamma}$ point of the 2$\times$2$\times$1 supercell, as the $\mathrm{M}$ point of the unit cell unfolds at the $\mathrm{\Gamma}$ point of the 2$\times$2$\times$1 supercell. Inspired by the eigen vectors of the softest phonon mode, we have applied in-plane and out-of-plane displacements to the $\mathrm{Zr}$ atoms with systematically varying amplitudes. $\mathrm{Se}$ and $\mathrm{Te}$ atoms have been fully relaxed using selective dynamics for each distorted configuration, and the total energies of those configurations have been computed. Figure \hyperref[fig:3]{3(a)} illustrates total energies that have been calculated for different distortion amplitudes ($\mathrm{\delta Zr}$). The energy profile clearly demonstrates a double-well potential, which indicates that the high-symmetry undistorted phase is unstable at low temperatures (below $\mathrm{T_{CDW}}$) and is prone to a CDW distortion \cite{lian2022intrinsic,yin2020theoretical}. In the distorted phase, the system resides in a bound state associated with one of the two symmetry-related minima, which denote energetically equal CDW patterns with opposite distortion phases. The significant deviation from harmonic (parabola) behavior and the emergence of distinct minima indicate substantial anharmonic effects, which are crucial for stabilizing the CDW ground state \cite{lian2022intrinsic}. At high temperature (above $\mathrm{T_{CDW}}$), the accessible thermal energy enables the system to surmount the double-well potential barrier effectively and fluctuates dynamically between distorted configurations, ultimately leading to the average restoration of the high-symmetry 1$\times$1$\times$1 structure \cite{zhang2023emergent}. The minima of the double-well potential for the monolayer $\mathrm{ZrSeTe}$ is located at $\mathrm{\delta Zr \, = 0.074 \, \AA}$ with about 0.40 meV/f.u. energy gain. This CDW energy gain is substantially lower than that of monolayer $\mathrm{ZrTe_2}$ \cite{zhang2023emergent}, suggesting that the system is driven away from the CDW-favored regime due to replacing one chalcogen layer of $\mathrm{Te}$ by the $\mathrm{Se}$ atom.

\begin{figure}[h]
\includegraphics[width=1\linewidth]{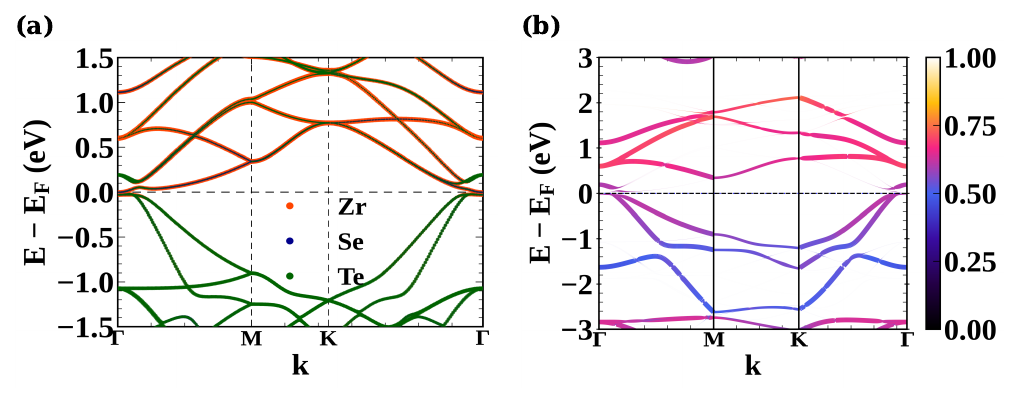}
\caption{\label{fig:4} (a) Electronic band structure of distorted 2$\times$2$\times$1 CDW phase of $\mathrm{ZrSeTe}$; (b) Corresponding band structure unfolded from 2$\times$2$\times$1 supercell to the primitive 1$\times$1$\times$1 Brillouin zone. SOC has not been considered in either calculation.}
\end{figure}

To obtain the CDW ground state, we have used the distorted structure corresponding to the minima of the double-well potential. Crystal structure of the distorted 2$
\times$2$\times$1 supercell of $\mathrm{ZrSeTe}$ has been displayed in the Fig. \hyperref[fig:3]{3(b)}. A detailed comparison between the undistorted and distorted crystal structure of a 2$\times$2$\times$1 supercell of $\mathrm{ZrSeTe}$ has been shown in Table \hyperref[tab:1]{1}. The distorted structure exhibits minor deviations in the lattice parameters and lattice angles from the ideal hexagonal symmetry, as well as subtle changes in $\mathrm{Zr-Se}$ and $\mathrm{Zr-Te}$ bond lengths. These results suggest a weak lattice distortion associated with the 2$\times$2$\times$1 supercell, consistent with the formation of a CDW instability. The crystal symmetry is reduced from p3m1 to p1 as a result of this distortion. It indicates the complete elimination of the rotational and mirror symmetries present in the undistorted phase. Figure \hyperref[fig:3]{3(c)} illustrates the phonon dispersion of the CDW ground state of $\mathrm{ZrSeTe}$, shown in blue. In contrast, the three unstable phonon modes of the undistorted 2$\times$2$\times$1 supercell are highlighted in orange. The absence of any significant imaginary phonon frequency ensures the dynamical stability of the CDW distorted configuration in the ground state. In the absence of soft phonon modes, the phonon spectrum is now independent of electronic temperature (electronic smearing here). The electronic band structure of the CDW distorted phase of $\mathrm{ZrSeTe}$ has been shown in Fig. \hyperref[fig:4]{4(a)}. In the band structure plot of Fig. \hyperref[fig:4]{4(a)}, several additional bands appear due to the band folding associated with the supercell periodicity. To separate the bands of the primitive cell from these folded replicas, the spectral weight has been evaluated from the overlap between the Bloch states of the supercell and those of the primitive cell using the band-unfolding technique \cite{Herath2020107080, Lang2024109063}. The resulting unfolded band structure, weighted by the spectral intensity, is presented in Fig. \hyperref[fig:4]{4(b)}. As a consequence of the lattice distortion, the $\mathrm{Zr}$ \textit{d} orbital dominated conduction band at the $\mathrm{M}$ point moves away from the Fermi level, while the $\mathrm{Te}$ \textit{p} orbital dominated valence bands around the $\mathrm{\Gamma}$ point move slightly towards the valence band. This reconstruction of the electronic structure leads to the opening of a small indirect band gap of approximately 20 meV between the conduction band minimum and valence band maximum. These results indicate that the CDW distortion drives $\mathrm{ZrSeTe}$ toward a low-band-gap semiconductor.

\begin{figure*}[t]
\includegraphics[width=0.75\linewidth]{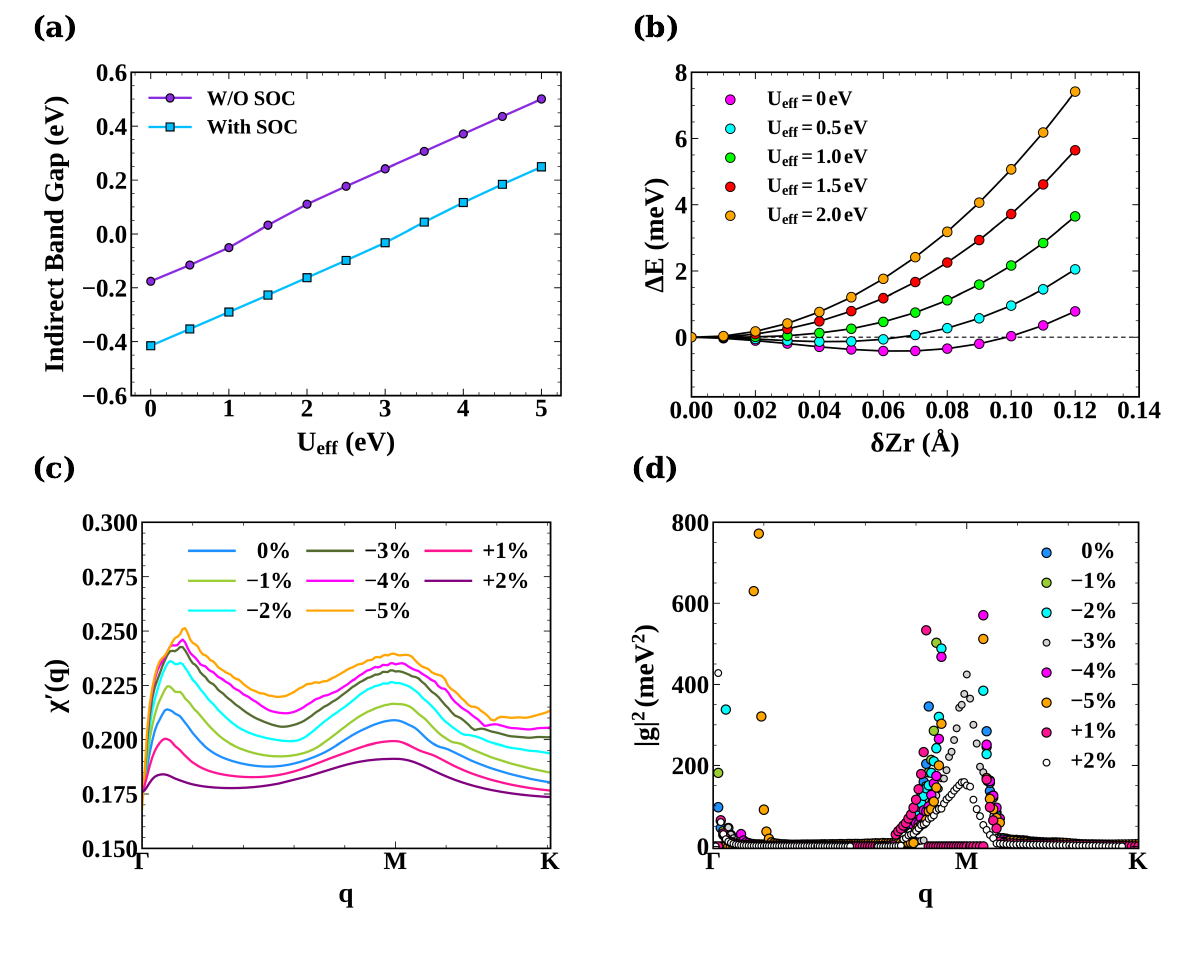}
\caption{\label{fig:5} (a) Indirect band gap of 1 T $\mathrm{ZrSeTe}$ monolayer vs Hubbard $\mathrm{U_{eff}}$; (b) CDW energy gain vs Hubbard $\mathrm{U_{eff}}$; (c) Real part of Lindhard total charge susceptibility along $\mathrm{\Gamma-M-K}$ path of IBZ for different strain; (d) Squared electron-phonon matrix element along $\mathrm{\Gamma-M-K}$ path of IBZ for different strain. SOC has not been considered for calculations in Figs. \hyperref[fig:5]{5(b-d)}.}
\end{figure*}

\subsection{\label{sec:3.4}Correlation induced changes in the electronic properties}

The electronic structure of systems with localized electrons can be frequently explained within the Hubbard framework, where the competition between the on-site Coulomb repulsion ($\mathrm{U}$) and the hopping amplitude ($\mathrm{t}$) controls the degree of localization \cite{pakdel2025effect}. Charge localization results in a Mott insulating state with a correlation-induced gap in the large $\mathrm{U/t}$ limit at half filling. This localization is frequently underestimated by conventional density functional theory (DFT), leading to inaccurate predictions of metallic or magnetic behavior. The $\mathrm{DFT+U}$ approach, which incorporates an on-site Coulomb correction, improves the description of localized states and is particularly useful for systems with lower bandwidth and reduced screening, where electron-electron interactions become significant \cite{imada1998metal}. Within the $\mathrm{DFT+U}$ paradigm, the on-site Coulomb interaction is often expressed in terms of an effective parameter, $\mathrm{U_{eff} = U - J}$, where $\mathrm{J}$ is Hund's exchange interaction. Specifically, in the Dudarev formulation \cite{dudarev1998electron}, the correction depends solely on $\mathrm{U_{eff}}$. The electronic band structures for varying $\mathrm{U_{eff}}$ values, with and without SOC, are presented in panels (a, c, e) and (b, d, f) respectively of Fig. S8 in the supplemental material \cite{supplementary}. As $\mathrm{U_{eff}}$ increases, the conduction bands touching the Fermi level at $\mathrm{M}$ point and the valence bands touching the Fermi level near $\mathrm{\Gamma}$ point shift towards the conduction and valence bands respectively. Consequently, an indirect band gap opens gradually with an increase of $\mathrm{U_{eff}}$. The variation of indirect gap with $\mathrm{U_{eff}}$ has been illustrated in the Fig. \hyperref[fig:5]{5(a)}. For $\mathrm{U_{eff}}$ values greater than 1 eV in non-SOC case and 3 eV in SOC case, 1T $\mathrm{ZrSeTe}$ gradually transitions from a semimetallic to semiconducting state. The change in the energy gain of the CDW state for different $\mathrm{U_{eff}}$ values has been displayed in Fig. \hyperref[fig:5]{5(b)}. SOC is neglected in this analysis. We observe that the energy gain of the CDW state decreases systematically with increasing $\mathrm{U_{eff}}$, indicating a progressive weakening of the instability. The CDW state eventually disappears as a result of the transition of 1T $\mathrm{ZrSeTe}$ monolayer from the semimetallic to the semiconducting state for $\mathrm{U_{eff}}$ values exceeding 1 eV.

\subsection{\label{sec:3.5}Impact of biaxial strain on CDW instability}

The weak nature of the CDW distortion in $\mathrm{ZrSeTe}$ suggests that its CDW instability may be strongly affected by lattice perturbations. Biaxial strain can significantly alter interatomic distances and electronic bandwidths, thereby affecting the CDW formation \cite{wei2017manipulating}. It motivates us to conduct a systematic investigation of the effect of biaxial compressive and tensile strains on the high-symmetry undistorted phase of 1T $\mathrm{ZrSeTe}$. Starting from the normal 1$\times$1$\times$1 undistorted structure, biaxial strains up to 5\% were quasistatically applied, and the corresponding phonon spectra were computed using a fixed electronic smearing of 0.005 Ry to examine possible strain-induced phonon softening and lattice instabilities. Figures S2 and S3 of the supplemental material \cite{supplementary} highlight the changes in electronic band structure and Fermi surface of the $\mathrm{ZrSeTe}$ monolayer under biaxial strains in the absence of SOC. Compressive strain causes the valence bands that intersect the Fermi level around the $\mathrm{\Gamma}$ point to shift towards the conduction band progressively. The slope of the linearly dispersive conduction band along the $\mathrm{M-K}$ path also increases. As a result, $\mathrm{ZrSeTe}$ exhibits more metallic character due to the larger band overlap under compressive strain. The dimensions of the electron pocket at the $\mathrm{M}$ point and the hole pockets surrounding the $\mathrm{\Gamma}$ point gradually expand under compressive strain. On the other hand, tensile strain induces a gradual shift of the valence bands intersecting the Fermi level around the $\mathrm{\Gamma}$ point towards the valence band. Linearly dispersive conduction band along $\mathrm{M-K}$ path flattens as tensile strain increases. The electron pocket at the $\mathrm{M}$ point disappears from the Fermi surface at 1\% tensile strain, indicating a Lifshitz transition \cite{feng2022superconductivity,wang2024theoretical}. As the tensile strain increases further, the system undergoes a metal-semiconductor transition at 3\%. For strains beyond this value, the band gap of the semiconducting phase increases monotonically. Figure S4 of the supplemental material \cite{supplementary} illustrates the phonon dispersion for various strain values in the absence of SOC. Under compressive strain, the area of imaginary phonon frequencies changes slightly relative to ambient conditions, although the CDW wave vector remains constant. This suggests that the CDW instability is predominantly resilient to compressive strain, except for 3\%. At this specific strain, the frequency of the soft phonon mode becomes positive, indicating a significant weakening of the CDW instability. On the other hand, under tensile strain, the CDW instability remains robust at 1\%. However, with a further increase in tensile strain, the instability gradually weakens and eventually disappears. To elucidate this behavior, we computed the real part of the Lindhard susceptibility and the squared EPC matrix element along the $\mathrm{\Gamma-M-K}$ path, as depicted in Figs. \hyperref[fig:5]{5(c)} and \hyperref[fig:5]{5(d)}. Figure \hyperref[fig:5]{5(c)} illustrates that the real part of susceptibility retains its local maximum at $\mathrm{M}$ point for all compressive strains and tensile strain up to 2\%, although its value varies with strain. This indicates that the tendency of electronic instability associated with the $\mathrm{M}$ point persists even under applied strain. Conversely, the EPC strength exhibits a more pronounced strain dependence. The squared EPC matrix element displays a divergence behavior around the $\mathrm{M}$ point for 1\%, 2\%, 4\%, and 5\% compressive strain, as well as for 1\% tensile strain, as illustrated in Fig. \hyperref[fig:5]{5(d)}. Therefore, it is consistent with the presence of a 2$\times$2$\times$1 CDW instability at those values of strain. However, at 3\% compressive strain and 2\% tensile strain, the squared EPC matrix element is significantly reduced and shows only a finite peak at the $\mathrm{M}$ point. This reduction in EPC lowers $\mathrm{\chi_{q,\nu}}$ in Eq. \hyperref[eq:6]{6}, thereby weakening the phonon softening as well as the CDW instability at those strains. The decrease in EPC at these strain values may be the result of strain-induced changes in orbital hybridization near the Fermi level or variations in the self-consistent potential associated with the phonon mode, as Eq. \hyperref[eq:9]{9} suggests. Furthermore, for tensile strains exceeding 2\%, the Fermi surface of $\mathrm{ZrSeTe}$ disappears, as shown in Figs. S2 and S3 of the supplemental material \cite{supplementary}. The absence of a Fermi surface suppresses the electronic instability and significantly weakens the EPC, ultimately leading to the gradual disappearance of the CDW instability at higher tensile strain.
\\
\par The evolution of electronic band structure, Fermi surface, and phonon dispersion in the presence of SOC under varying strain values is illustrated in Figs. S5, S6, and S7 of the supplemental material \cite{supplementary}. At 1\% compressive strain, two additional bands, each spin-split due to SOC, intersect the Fermi level, indicating a Lifshitz transition \cite{feng2022superconductivity,wang2024theoretical}. As compressive strain increases, the overall changes in the band structure and the size of the Fermi sheets follow a trend similar to that observed in the non-SOC calculations. The incorporation of SOC induces a metal-semiconductor transition at 5\% tensile strain. The phonon dispersion calculations reveal that the instability increases slightly at 1\% compressive strain, while additional compressive strain progressively weakens the instability. Conversely, tensile strain reduces the size of the Fermi sheets and gradually suppresses the electronic instability, eventually leading to the complete disappearance of the CDW phase. The incorporation of SOC may modify the band hybridization or the self-consistent potential, so affecting the EPC strength and, subsequently, the stability of the CDW phase.

\begin{figure*}[t]
\includegraphics[width=1\linewidth]{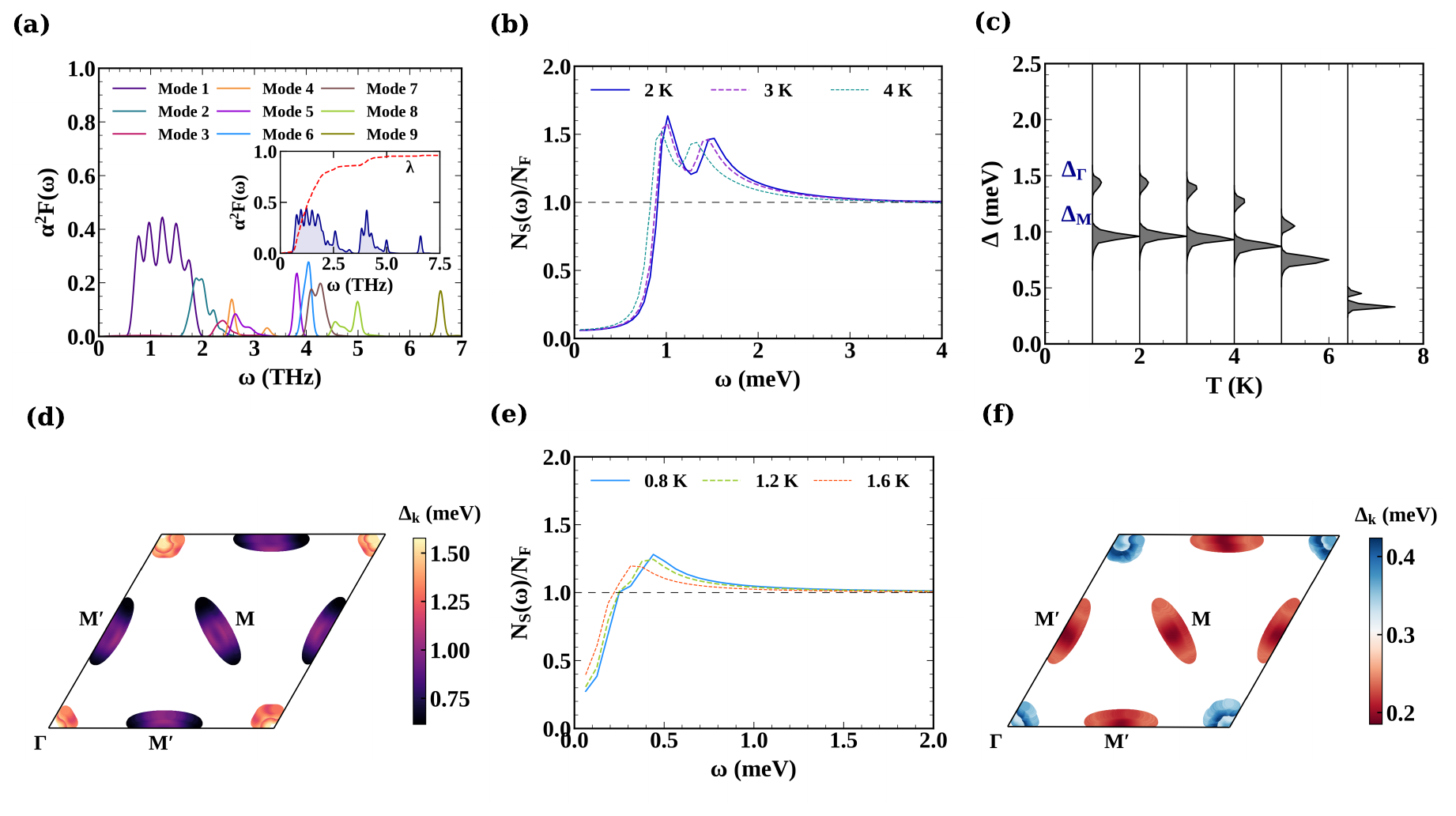}
\caption{\label{fig:6} (a) Main: Modewise isotropic Eliashberg spectral function $\mathrm{\alpha^2F(\omega)}$ as a function of frequency, Inset: total $\mathrm{\alpha^2F(\omega)}$ and cumulative EPC strength $\mathrm{\lambda}$ vs frequency; Normalized quasiparticle density of states in the superconducting state as a function of frequency at three temperatures for (b) non-SOC case, (d) SOC case. Here, the dashed black line denotes the density of states in the normal state, set to 1 at the Fermi level; (c) the variation of anisotropic superconducting gap on the Fermi surface of undistorted $\mathrm{ZrSeTe}$ monolayer with temperature; Color plot of momentum resolved superconducting gap, $\mathrm{\Delta_k}$ on the electronic states within the Fermi window for (d) non-SOC case, (f) SOC case. Here, SOC has been neglected in plots of (a) and (c).}
\end{figure*}

\subsection{\label{sec:3.6}Superconductivity of the undistorted phase}

We have employed the Migdal-Eliashberg approach \cite{migdal1958interaction,eliashberg1960interactions}, and the McMillan modified Allen-Dynes empirical \cite{allen1975transition,mcmillan1968transition} formula to compute the superconductivity of the high-temperature undistorted phase. A marginally increased smearing width has been employed to render the soft mode frequency positive, enabling its involvement in superconductivity. Electronic states at the Fermi level are primarily characterized by $\mathrm{Te}$ $\textit{p}$ orbitals and $\mathrm{Zr}$ $\textit{d}$ orbitals. Phonon modes that participate in electron-phonon coupling have been displayed in the phonon dispersion plot of Fig. \hyperref[fig:2]{2(a)}. In the high temperature phase of the $\mathrm{ZrSeTe}$ monolayer, the superconductivity emerges as a result of the robust coupling of those electronic states with the phonon mode exhibiting the Kohn anomaly \cite{yin2020theoretical} and other phonon modes shown in Fig. \hyperref[fig:2]{2(a)}. The computed isotropic Eliashberg spectral function and cumulative EPC strength for the undistorted phase of the ZrSeTe monolayer in the absence of SOC are presented in Fig. \hyperref[fig:6]{6(a)}. The Eliashberg spectral function of the $\mathrm{ZrSeTe}$ monolayer comprises three clusters. Notably, mode 1, which exhibits a Kohn anomaly \cite{yin2020theoretical}, predominantly contributes to $\mathrm{\alpha^2F(\omega)}$ in the low-frequency domain. The value of the cumulative EPC strength ($\mathrm{\lambda}$) is about 0.96. The solutions of the Migdal-Eliashberg equations (Eqs. \hyperref[eq:4]{4},\hyperref[eq:5]{5}) \cite{migdal1958interaction,eliashberg1960interactions} along the imaginary energy axis are analytically continued to the real energy axis using Pad\'e approximation. The quasiparticle density of states is then retrieved from the poles of the normal Green's function using the EPW code \cite{ponce2016epw,margine2013anisotropic,giustino2007electron}. Figure \hyperref[fig:6]{6(b)} illustrates the quasiparticle density of states of the $\mathrm{ZrSeTe}$ monolayer in superconducting states without consideration of SOC. The presence of two peaks in this plot indicates two superconducting gaps in the undistorted phase of the $\mathrm{ZrSeTe}$ monolayer. The two-gap superconductivity in the absence of SOC is also apparent from the plot of the superconducting gap on the Fermi surface as a function of temperature, illustrated in Fig. \hyperref[fig:6]{6(c)}. The larger gap originates from the valence bands intersecting the Fermi level near the $\mathrm{\Gamma}$ point, whereas the smaller gap arises from the conduction band crossing the Fermi level at the $\mathrm{M}$ point. Both superconducting gaps vanish around the transition temperature, $\mathrm{T_c^{Aniso}}$ = 6.46 K (See Fig. \hyperref[fig:6]{6(c)}). Nevertheless, its superconducting critical temperature ($\mathrm{T_c^{MAD}}$), in the absence of SOC, has been determined to be 4.25 K via the Allen-Dynes modified McMillan formula \cite{allen1975transition,mcmillan1968transition}. This clearly indicates the occurrence of anisotropic superconductivity in the undistorted phase of the $\mathrm{ZrSeTe}$ monolayer. Table \hyperref[tab:2]{II} summarizes the superconducting transition temperature ($\mathrm{T_c}$) of the undistorted $\mathrm{ZrSeTe}$ monolayer at low, intermediate, and large values of effective Coulomb potential ($\mathrm{\mu_c^*}$).

\begin{table}[h]
\caption{\label{tab:2}%
Variation of superconducting transition temperature ($\mathrm{T_c}$) of the undistorted $\mathrm{ZrSeTe}$ monolayer with effective Coulomb potential ($\mathrm{\mu_c^\ast}$). SOC has not been considered here.
}
\renewcommand{\arraystretch}{1.25}     
\setlength{\tabcolsep}{10pt} 
\begin{ruledtabular}
\begin{tabular}{ccc}
$\mathrm{\mu_c^\ast}$ & $\mathrm{T_c^{MAD}}$ (K) & $\mathrm{T_c^{Aniso}}$ (K)\\
\colrule
0.05 & 5.35 & 7.28\\
0.10 & 4.25 & 6.46\\
0.15 & 3.21 & 5.73\\
\end{tabular}
\end{ruledtabular}
\end{table}

An elevation in $\mathrm{\mu_c^*}$ results in a decrease in $\mathrm{T_c}$ of $\mathrm{ZrSeTe}$. Figure \hyperref[fig:6]{6(d)} depicts the color plot of the momentum-resolved superconducting gap on the electronic states within the Fermi window surrounding the Fermi level. The two-gap nature and the gap anisotropy are clearly seen in this plot. The quasiparticle density of states in the superconducting states has been illustrated in Fig. \hyperref[fig:6]{6(e)} in the presence of SOC. It still displays two peaks, but they are at lower frequencies than those in the non-SOC case. In the presence of SOC, the shift of the conduction band at $\mathrm{M}$ point away from the Fermi level and the lifting of the valence band degeneracy at $\mathrm{\Gamma}$ point (Fig. S1(b) of the supplemental material \cite{supplementary}) reduce the number of electronic states available for scattering. This leads to the suppression of both the critical temperature and the superconducting gap. In the presence of SOC, the calculated cumulative EPC strength is 0.90, and the critical temperatures, determined using the Allen-Dyne modified McMillan formula \cite{allen1975transition,mcmillan1968transition} and the Migdal-Eliashberg equation \cite{migdal1958interaction,eliashberg1960interactions}, are 2.16 K and 1.96 K, respectively. The color plot of the superconducting gap values on the electronic states within the Fermi window, attached in Fig. \hyperref[fig:6]{6(f)}, shows a reduced separation between the two gap magnitudes. Moreover, the presence of SOC suppresses the gap anisotropy.

\section{\label{sec:4}Conclusions}

In summary, a first-principles investigation of the CDW instability and superconductivity in 1T $\mathrm{ZrSeTe}$ Janus monolayer reveals a strong softening of the acoustic phonon mode at the $\mathrm{M}$ point in phonon dispersion. This strongly supports a 2$\times$2$\times$1 CDW instability like its non-Janus counterpart $\mathrm{ZrTe_2}$ monolayer. A pronounced peak in the EPC matrix element, together with an enhancement of the real part of the static susceptibility, arising from both intra- and interband contributions, leads to the emergence of a Kohn anomaly at the M point of the Brillouin zone of $\mathrm{ZrSeTe}$. Replacing one $\mathrm{Te}$ layer by $\mathrm{Se}$ militates against the CDW-instability. The CDW distortion reconstructs the electronic structure and drives the system from a semi-metallic phase to a semiconducting phase. On incorporating a local correlation (Hubbard $\mathrm{U_{eff}}$) we observe that an increasing $\mathrm{U_{eff}}$ progressively destabilizes the CDW instability by suppressing the semi-metallic Fermi surface. Beyond $\mathrm{U_{eff}}$ = 3 eV (with SOC), a correlation-induced gap precedes the CDW-induced gap at the Fermi level, eliminating the CDW instability. The semiconducting state in this case is correlation-driven, rather than due to an FS instability via phonons. To assess the robustness of the instability, biaxial strain has been applied quasi-statically to the high-symmetry undistorted 1$\times$1$\times$1 structure. The CDW instability remains robust at 1\% compressive strain and gradually weakens with further compression. Tensile strain, on the other hand, suppresses the instability and eventually drives the system toward an instability-free regime, accompanied by a metal-to-semiconductor transition. Our findings also provide evidence that SOC and local correlation are significant factors in the modulation of the CDW instability.
\\
\par We have also examined the superconducting properties of the $\mathrm{ZrSeTe}$ monolayer in its high temperature undistorted phase. Our analysis reveals that the strong coupling between the electronic bands crossing the Fermi level and the soft phonon mode at the $\mathrm{M}$ point provides the dominant contribution to superconductivity in this system. In the absence of the CDW phase, $\mathrm{ZrSeTe}$ is predicted to host a two-gap anisotropic superconducting state mediated by electron-phonon interaction. Furthermore, the inclusion of SOC significantly reduces both the superconducting gap and the transition temperature. Our findings establish the 1T $\mathrm{ZrSeTe}$ monolayer as a promising platform for investigating low-energy collective phenomena, particularly charge density wave order and superconductivity, in low-dimensional Janus materials.

\vspace*{3pt}
\begin{acknowledgments}
We acknowledge the National Supercomputing Mission (NSM) for providing computational resources through the PARAM Shakti high-performance computing facility at IIT Kharagpur. PARAM Shakti has been established by the Centre for Development of Advanced Computing (C-DAC) under the support of the Ministry of Electronics and Information Technology (MeitY) and the Department of Science and Technology (DST), Government of India. Subhajit Pramanick is thankful to Dr. Chao-Sheng Lian for useful discussions and also acknowledges IIT Kharagpur for the research fellowship.
\end{acknowledgments}

\nocite{*}
\bibliography{Manuscript}

@article{chang2012direct,
  title={Direct observation of competition between superconductivity and charge density wave order in $\mathrm{YBa_2Cu_3O_{6.67}}$},
  author={Chang, J and Blackburn, Elizabeth and Holmes, AT and Christensen, Niels B and Larsen, Jacob and Mesot, J and Liang, Ruixing and Bonn, DA and Hardy, WN and Watenphul, A and others},
  journal={Nature Physics},
  volume={8},
  number={12},
  pages={871--876},
  year={2012},
  publisher={Nature Publishing Group UK London},
  doi={10.1038/nphys2456},
  url={https://doi.org/10.1038/nphys2456}
}

@article{da2014ubiquitous,
  title={Ubiquitous interplay between charge ordering and high-temperature superconductivity in cuprates},
  author={da Silva Neto, Eduardo H and Aynajian, Pegor and Frano, Alex and Comin, Riccardo and Schierle, Enrico and Weschke, Eugen and Gyenis, Andr{\'a}s and Wen, Jinsheng and Schneeloch, John and Xu, Zhijun and others},
  journal={Science},
  volume={343},
  number={6169},
  pages={393--396},
  year={2014},
  publisher={American Association for the Advancement of Science},
  doi={10.1126/science.1243479},
  url={https://www.science.org/doi/abs/10.1126/science.1243479}
}

@article{wilson1974charge,
  title={{Charge-Density Waves in Metallic, Layered, Transition-Metal Dichalcogenides}},
  author={Wilson, J. A. and Di Salvo, F. J. and Mahajan, S.},
  journal={Phys. Rev. Lett.},
  volume={32},
  issue={16},
  pages={882--885},
  numpages={0},
  year={1974},
  month={Apr},
  publisher={American Physical Society},
  doi={10.1103/PhysRevLett.32.882},
  url={https://link.aps.org/doi/10.1103/PhysRevLett.32.882}
}

@article{wilson1975charge,
  author={J.A. Wilson and F.J. Di Salvo and S. Mahajan},
  title={Charge-density waves and superlattices in the metallic layered transition metal dichalcogenides},
  journal={Advances in Physics},
  volume={24},
  number={2},
  pages={117--201},
  year={1975},
  publisher={Taylor \& Francis},
  doi={10.1080/00018737500101391},
  url={https://doi.org/10.1080/00018737500101391}
}

@article{halperin1968possible,
  title={{Possible Anomalies at a Semimetal-Semiconductor Transistion}},
  author={Halperin, BI and Rice, TM},
  journal={Rev. Mod. Phys.},
  volume={40},
  issue={4},
  pages={755--766},
  numpages={0},
  year={1968},
  month={Oct},
  publisher={American Physical Society},
  doi={10.1103/RevModPhys.40.755},
  url={https://link.aps.org/doi/10.1103/RevModPhys.40.755}
}

@incollection{halperin1968excitonic,
  title={The excitonic state at the semiconductor-semimetal transition},
  author={Halperin, BI and Rice, TM},
  booktitle={Solid State Physics},
  volume={21},
  pages={115--192},
  year={1968},
  publisher={Elsevier},
  doi={https://doi.org/10.1016/S0081-1947(08)60740-7},
  url={https://www.sciencedirect.com/science/article/pii/S0081194708607407}
}

@article{lian2022intrinsic,
  title={Intrinsic and doping-enhanced superconductivity in monolayer {$1H\text{\ensuremath{-}}{\mathrm{TaS}}_{2}$}: Critical role of charge ordering and spin-orbit coupling},
  author={Lian, Chao-Sheng and Heil, Christoph and Liu, Xiaoyu and Si, Chen and Giustino, Feliciano and Duan, Wenhui},
  journal={Phys. Rev. B},
  volume={105},
  issue={18},
  pages={L180505},
  numpages={6},
  year={2022},
  month={May},
  publisher={American Physical Society},
  doi={10.1103/PhysRevB.105.L180505},
  url={https://link.aps.org/doi/10.1103/PhysRevB.105.L180505}
}

@article{zheng2019electron,
  title={Electron-phonon coupling and the coexistence of superconductivity and charge-density wave in monolayer {${\mathrm{NbSe}}_{2}$}},
  author={Zheng, Feipeng and Feng, Ji},
  journal={Phys. Rev. B},
  volume={99},
  issue={16},
  pages={161119},
  numpages={5},
  year={2019},
  month={Apr},
  publisher={American Physical Society},
  doi={10.1103/PhysRevB.99.161119},
  url={https://link.aps.org/doi/10.1103/PhysRevB.99.161119}
}

@article{zhang2023emergent,
  title={Emergent charge density wave order in the monolayer limit of {$1T\text{\ensuremath{-}}{\mathrm{TiTe}}_{2}$} and {$1T\text{\ensuremath{-}}{\mathrm{ZrTe}}_{2}$}},
  author={Zhang, Jiayuan and Wang, Fei and Lian, Chao-Sheng},
  journal={Phys. Rev. B},
  volume={108},
  issue={16},
  pages={165421},
  numpages={10},
  year={2023},
  month={Oct},
  publisher={American Physical Society},
  doi={10.1103/PhysRevB.108.165421},
  url={https://link.aps.org/doi/10.1103/PhysRevB.108.165421}
}

@article{shao2016manipulating,
  title={Manipulating charge density waves in {$1T\text{\ensuremath{-}}{\mathrm{TaS}}_{2}$} by charge-carrier doping: A first-principles investigation},
  author={Shao, D. F. and Xiao, R. C. and Lu, W. J. and Lv, H. Y. and Li, J. Y. and Zhu, X. B. and Sun, Y. P.},
  journal={Phys. Rev. B},
  volume={94},
  issue={12},
  pages={125126},
  numpages={9},
  year={2016},
  month={Sep},
  publisher={American Physical Society},
  doi={10.1103/PhysRevB.94.125126},
  url={https://link.aps.org/doi/10.1103/PhysRevB.94.125126}
}

@article{wei2017manipulating,
  title={Manipulating charge density wave order in monolayer {$1T\ensuremath{-}{\mathrm{TiSe}}_{2}$} by strain and charge doping: A first-principles investigation},
  author={Wei, M. J. and Lu, W. J. and Xiao, R. C. and Lv, H. Y. and Tong, P. and Song, W. H. and Sun, Y. P.},
  journal={Phys. Rev. B},
  volume={96},
  issue={16},
  pages={165404},
  numpages={6},
  year={2017},
  month={Oct},
  publisher={American Physical Society},
  doi={10.1103/PhysRevB.96.165404},
  url={https://link.aps.org/doi/10.1103/PhysRevB.96.165404}
}

@article{ren2022semiconductor,
  title={{Semiconductor--Metal Phase Transition and Emergent Charge Density Waves in 1 T-$\mathrm{ZrX_2}$ (X= Se, Te) at the Two-Dimensional Limit}},
  author={Ren, Ming-Qiang and Han, Sha and Fan, Jia-Qi and Wang, Li and Wang, Pengdong and Ren, Wei and Peng, Kun and Li, Shujing and Wang, Shu-Ze and Zheng, Fa-Wei and others},
  journal={Nano Letters},
  volume={22},
  number={1},
  pages={476--484},
  year={2022},
  publisher={ACS Publications},
  doi={10.1021/acs.nanolett.1c04372},
  url={https://doi.org/10.1021/acs.nanolett.1c04372}
}

@article{patel2024electron,
  title={{Electron-phonon coupling, critical temperatures, and gaps in ${\mathrm{NbSe}}_{2}/{\mathrm{MoS}}_{2}$ Ising superconductors}},
  author={Patel, Shubham and Jena, Soumyasree and Taraphder, A.},
  journal={Phys. Rev. B},
  volume={110},
  issue={1},
  pages={014507},
  numpages={8},
  year={2024},
  month={Jul},
  publisher={American Physical Society},
  doi={10.1103/PhysRevB.110.014507},
  url={https://link.aps.org/doi/10.1103/PhysRevB.110.014507}
}

@article{tan2021charge,
  title={{Charge Density Waves and Electronic Properties of Superconducting Kagome Metals}},
  author={Tan, Hengxin and Liu, Yizhou and Wang, Ziqiang and Yan, Binghai},
  journal={Phys. Rev. Lett.},
  volume={127},
  issue={4},
  pages={046401},
  numpages={6},
  year={2021},
  month={Jul},
  publisher={American Physical Society},
  doi={10.1103/PhysRevLett.127.046401},
  url={https://link.aps.org/doi/10.1103/PhysRevLett.127.046401}
}

@article{lin2022multidome,
  title={Multidome superconductivity in charge density wave kagome metals},
  author={Lin, Yu-Ping and Nandkishore, Rahul M.},
  journal={Phys. Rev. B},
  volume={106},
  issue={6},
  pages={L060507},
  numpages={8},
  year={2022},
  month={Aug},
  publisher={American Physical Society},
  doi={10.1103/PhysRevB.106.L060507},
  url={https://link.aps.org/doi/10.1103/PhysRevB.106.L060507}
}

@article{cao2023competing,
  title={Competing charge-density wave instabilities in the kagome metal $\mathrm{ScV_6Sn_6}$},
  author={Cao, Saizheng and Xu, Chenchao and Fukui, Hiroshi and Manjo, Taishun and Dong, Ying and Shi, Ming and Liu, Yang and Cao, Chao and Song, Yu},
  journal={Nature Communications},
  volume={14},
  number={1},
  pages={7671},
  year={2023},
  publisher={Nature Publishing Group UK London},
  doi={10.1038/s41467-023-43454-1},
  url={https://doi.org/10.1038/s41467-023-43454-1}
}

@book{peierls1955quantum,
  title={Quantum theory of solids},
  author={Peierls, Rudolf Ernst},
  year={1955},
  publisher={Oxford university press}
}

@article{kohn1959image,
  title={{Image of the Fermi Surface in the Vibration Spectrum of a Metal}},
  author={Kohn, W.},
  journal={Phys. Rev. Lett.},
  volume={2},
  issue={9},
  pages={393--394},
  numpages={0},
  year={1959},
  month={May},
  publisher={American Physical Society},
  doi={10.1103/PhysRevLett.2.393},
  url={https://link.aps.org/doi/10.1103/PhysRevLett.2.393}
}

@article{zhu2015classification,
  title={Classification of charge density waves based on their nature},
  author={Zhu, Xuetao and Cao, Yanwei and Zhang, Jiandi and Plummer, EW and Guo, Jiandong},
  journal={Proceedings of the National Academy of Sciences},
  volume={112},
  number={8},
  pages={2367--2371},
  year={2015},
  publisher={National Academy of Sciences},
  doi={10.1073/pnas.1424791112},
  url={https://www.pnas.org/doi/abs/10.1073/pnas.1424791112}
}

@article{kaboudvand2022fermi,
  title={{Fermi surface nesting and the Lindhard response function in the kagome superconductor $\mathrm{CsV_3Sb_5}$}},
  author={Kaboudvand, Farnaz and Teicher, Samuel ML and Wilson, Stephen D and Seshadri, Ram and Johannes, Michelle D},
  journal={Applied Physics Letters},
  volume={120},
  number={11},
  year={2022},
  publisher={AIP Publishing},
  url={https://doi.org/10.1063/5.0081081}
}

@article{pramanick2025pressureinducedevolutionanisotropic,
  title={{Pressure induced evolution of anisotropic superconductivity and Fermi surface nesting in a ternary boride}},
  author={Pramanick, Subhajit and Chakraborty, Sudip and Taraphder, A.},
  journal={Phys. Rev. B},
  volume={112},
  issue={21},
  pages={214517},
  numpages={13},
  year={2025},
  month={Dec},
  publisher={American Physical Society},
  doi={10.1103/b627-nvxl},
  url={https://link.aps.org/doi/10.1103/b627-nvxl}
}

@article{jerome1967excitonic,
  title={{Excitonic Insulator}},
  author={J\'erome, D. and Rice, T. M. and Kohn, W.},
  journal={Phys. Rev.},
  volume={158},
  issue={2},
  pages={462--475},
  numpages={0},
  year={1967},
  month={Jun},
  publisher={American Physical Society},
  doi={10.1103/PhysRev.158.462},
  url={https://link.aps.org/doi/10.1103/PhysRev.158.462}
}

@article{taraphder2011preformed,
  title={{Preformed Excitonic Liquid Route to a Charge Density Wave in $2H\mathrm{\text{\ensuremath{-}}}{\mathrm{TaSe}}_{2}$}},
  author={Taraphder, A. and Koley, S. and Vidhyadhiraja, N. S. and Laad, M. S.},
  journal={Phys. Rev. Lett.},
  volume={106},
  issue={23},
  pages={236405},
  numpages={4},
  year={2011},
  month={Jun},
  publisher={American Physical Society},
  doi={10.1103/PhysRevLett.106.236405},
  url={https://link.aps.org/doi/10.1103/PhysRevLett.106.236405}
}

@article{koley2015unusual,
  title={{The unusual normal state and charge-density-wave order in 2H-NbSe$_2$}},
  author={Koley, S and Mohanta, N and Taraphder, A},
  journal={Journal of Physics: Condensed Matter},
  volume={27},
  number={18},
  pages={185601},
  year={2015},
  publisher={IOP Publishing},
  doi={10.1088/0953-8984/27/18/185601},
  url={https://doi.org/10.1088/0953-8984/27/18/185601}
}

@article{koley2020charge,
  title={Charge density wave and superconductivity in transition metal dichalcogenides},
  author={Koley, Sudipta and Mohanta, Narayan and Taraphder, Arghya},
  journal={The European Physical Journal B},
  volume={93},
  number={5},
  pages={77},
  year={2020},
  publisher={Springer},
  doi={10.1140/epjb/e2020-100522-5},
  url={https://doi.org/10.1140/epjb/e2020-100522-5}
}

@article{cercellier2007evidence,
  title={{Evidence for an Excitonic Insulator Phase in $1T\mathrm{\text{\ensuremath{-}}}{\mathrm{TiSe}}_{2}$}},
  author={Cercellier, H. and Monney, C. and Clerc, F. and Battaglia, C. and Despont, L. and Garnier, M. G. and Beck, H. and Aebi, P. and Patthey, L. and Berger, H. and Forr\'o, L.},
  journal={Phys. Rev. Lett.},
  volume={99},
  issue={14},
  pages={146403},
  numpages={4},
  year={2007},
  month={Oct},
  publisher={American Physical Society},
  doi={10.1103/PhysRevLett.99.146403},
  url={https://link.aps.org/doi/10.1103/PhysRevLett.99.146403}
}

@article{monney2011exciton,
  title={{Exciton Condensation Driving the Periodic Lattice Distortion of $1T\mathrm{\text{\ensuremath{-}}}{\mathrm{TiSe}}_{2}$}},
  author={Monney, C. and Battaglia, C. and Cercellier, H. and Aebi, P. and Beck, H.},
  journal={Phys. Rev. Lett.},
  volume={106},
  issue={10},
  pages={106404},
  numpages={4},
  year={2011},
  month={Mar},
  publisher={American Physical Society},
  doi={10.1103/PhysRevLett.106.106404},
  url={https://link.aps.org/doi/10.1103/PhysRevLett.106.106404}
}

@article{wegner2020evidence,
  title={{Evidence for pseudo--Jahn-Teller distortions in the charge density wave phase of $1T\text{\ensuremath{-}}{\mathrm{TiSe}}_{2}$}},
  author={Wegner, A. and Zhao, J. and Li, J. and Yang, J. and Anikin, A. A. and Karapetrov, G. and Esfarjani, K. and Louca, D. and Chatterjee, U.},
  journal={Phys. Rev. B},
  volume={101},
  issue={19},
  pages={195145},
  numpages={6},
  year={2020},
  month={May},
  publisher={American Physical Society},
  doi={10.1103/PhysRevB.101.195145},
  url={https://link.aps.org/doi/10.1103/PhysRevB.101.195145}
}

@article{wang2022origin,
  title={{Origin of charge density wave in the layered kagome metal ${\mathrm{CsV}}_{3}{\mathrm{Sb}}_{5}$}},
  author={Wang, Chongze and Liu, Shuyuan and Jeon, Hyunsoo and Cho, Jun-Hyung},
  journal={Phys. Rev. B},
  volume={105},
  issue={4},
  pages={045135},
  numpages={6},
  year={2022},
  month={Jan},
  publisher={American Physical Society},
  doi={10.1103/PhysRevB.105.045135},
  url={https://link.aps.org/doi/10.1103/PhysRevB.105.045135}
}

@article{yin2020theoretical,
  title={Theoretical investigation of charge density wave instability in {${\mathrm{CuS}}_{2}$}},
  author={Yin, Yuxin and Coulter, Jennifer and Ciccarino, Christopher J. and Narang, Prineha},
  journal={Phys. Rev. Mater.},
  volume={4},
  issue={10},
  pages={104001},
  numpages={7},
  year={2020},
  month={Oct},
  publisher={American Physical Society},
  doi={10.1103/PhysRevMaterials.4.104001},
  url={https://link.aps.org/doi/10.1103/PhysRevMaterials.4.104001}
}

@article{gabovich2002charge,
  title={{Charge-and spin-density waves in existing superconductors: competition between Cooper pairing and Peierls or excitonic instabilities}},
  author={Gabovich, AM and Voitenko, AI and Ausloos, Marcel},
  journal={Physics Reports},
  volume={367},
  number={6},
  pages={583--709},
  year={2002},
  publisher={Elsevier},
  doi={https://doi.org/10.1016/S0370-1573(02)00029-7},
  url={https://www.sciencedirect.com/science/article/pii/S0370157302000297}
}

@article{hoesch2009giant,
  title={{Giant Kohn Anomaly and the Phase Transition in Charge Density Wave ${\mathrm{ZrTe}}_{3}$}},
  author={Hoesch, Moritz and Bosak, Alexey and Chernyshov, Dmitry and Berger, Helmuth and Krisch, Michael},
  journal={Phys. Rev. Lett.},
  volume={102},
  issue={8},
  pages={086402},
  numpages={4},
  year={2009},
  month={Feb},
  publisher={American Physical Society},
  doi={10.1103/PhysRevLett.102.086402},
  url={https://link.aps.org/doi/10.1103/PhysRevLett.102.086402}
}

@article{monceau2012electronic,
  title={Electronic crystals: an experimental overview},
  author={Monceau, Pierre},
  journal={Advances in Physics},
  volume={61},
  number={4},
  pages={325--581},
  year={2012},
  publisher={Taylor \& Francis},
  doi={10.1080/00018732.2012.719674},
  url={https://doi.org/10.1080/00018732.2012.719674}
}

@article{wang2020band,
  title={{Band insulator to Mott insulator transition in 1 $\mathrm{T-TaS_2}$}},
  author={Wang, YD and Yao, WL and Xin, ZM and Han, TT and Wang, ZG and Chen, L and Cai, C and Li, Yuan and Zhang, Y},
  journal={Nature communications},
  volume={11},
  number={1},
  pages={4215},
  year={2020},
  publisher={Nature Publishing Group UK London},
  doi={10.1038/s41467-020-18040-4},
  url={https://doi.org/10.1038/s41467-020-18040-4}
}

@article{cho2017correlated,
  title={{Correlated electronic states at domain walls of a Mott-charge-density-wave insulator 1 $\mathrm{T-TaS_2}$}},
  author={Cho, Doohee and Gye, Gyeongcheol and Lee, Jinwon and Lee, Sung-Hoon and Wang, Lihai and Cheong, Sang-Wook and Yeom, Han Woong},
  journal={Nature communications},
  volume={8},
  number={1},
  pages={392},
  year={2017},
  publisher={Nature Publishing Group UK London},
  doi={10.1038/s41467-017-00438-2},
  url={https://doi.org/10.1038/s41467-017-00438-2}
}

@article{ma2016metallic,
  title={{A metallic mosaic phase and the origin of Mott-insulating state in $\mathrm{1T-TaS_2}$}},
  author={Ma, Liguo and Ye, Cun and Yu, Yijun and Lu, Xiu Fang and Niu, Xiaohai and Kim, Sejoong and Feng, Donglai and Tom{\'a}nek, David and Son, Young-Woo and Chen, Xian Hui and others},
  journal={Nature communications},
  volume={7},
  number={1},
  pages={10956},
  year={2016},
  publisher={Nature Publishing Group UK London},
  doi={10.1038/ncomms10956},
  url={https://doi.org/10.1038/ncomms10956}
}

@article{joshi2019short,
  title={{Short-range charge density wave order in $2H\text{\ensuremath{-}}\mathrm{T}\mathrm{a}{\mathrm{S}}_{2}$}},
  author={Joshi, Jaydeep and Hill, Heather M. and Chowdhury, Sugata and Malliakas, Christos D. and Tavazza, Francesca and Chatterjee, Utpal and Hight Walker, Angela R. and Vora, Patrick M.},
  journal={Phys. Rev. B},
  volume={99},
  issue={24},
  pages={245144},
  numpages={9},
  year={2019},
  month={Jun},
  publisher={American Physical Society},
  doi={10.1103/PhysRevB.99.245144},
  url={https://link.aps.org/doi/10.1103/PhysRevB.99.245144}
}

@article{wijayaratne2017spectroscopic,
  title={{Spectroscopic signature of moment-dependent electron--phonon coupling in 2 $\mathrm{H-TaS_2}$}},
  author={Wijayaratne, Kapila and Zhao, Junjing and Malliakas, Christos and Chung, Duck Young and Kanatzidis, Mercouri G and Chatterjee, Utpal},
  journal={Journal of Materials Chemistry C},
  volume={5},
  number={43},
  pages={11310--11316},
  year={2017},
  publisher={Royal Society of Chemistry},
  doi={10.1039/C7TC02641B},
  url={http://dx.doi.org/10.1039/C7TC02641B}
}

@article{wang2019evidence,
  title={Evidence of charge density wave with anisotropic gap in a monolayer {${\mathrm{VTe}}_{2}$} film},
  author={Wang, Yuan and Ren, Junhai and Li, Jiaheng and Wang, Yujia and Peng, Huining and Yu, Pu and Duan, Wenhui and Zhou, Shuyun},
  journal={Phys. Rev. B},
  volume={100},
  issue={24},
  pages={241404},
  numpages={6},
  year={2019},
  month={Dec},
  publisher={American Physical Society},
  doi={10.1103/PhysRevB.100.241404},
  url={https://link.aps.org/doi/10.1103/PhysRevB.100.241404}
}

@article{wang2025unraveling,
  title={Unraveling competing charge orders in monolayer {${\mathrm{VTe}}_{2}$}},
  author={Wang, Qiongru and Wang, Fei and Lian, Chao-Sheng},
  journal={Phys. Rev. B},
  volume={112},
  issue={24},
  pages={245410},
  numpages={9},
  year={2025},
  month={Dec},
  publisher={American Physical Society},
  doi={10.1103/ggfw-ypkp},
  url={https://link.aps.org/doi/10.1103/ggfw-ypkp}
}

@article{Sonika_2025,
  title={{Extended Kohler’s scaling and isosbestic point in the charge density wave state of $\mathrm{1T-VSe_2}$}},
  author={Sonika and Gangwar, Sunil and Kumar, Pankaj and Taraphder, A and Yadav, C S},
  journal={Journal of Physics: Condensed Matter},
  number={19},
  pages={195601},
  year={2025},
  publisher={IOP Publishing},
  doi = {10.1088/1361-648X/adc6e4},
  url = {https://doi.org/10.1088/1361-648X/adc6e4}
}

@article{chen2017emergence,
  title={{Emergence of charge density waves and a pseudogap in single-layer $\mathrm{TiTe_2}$}},
  author={Chen, Peng and Pai, Woei Wu and Chan, Y-H and Takayama, A and Xu, C-Z and Karn, A and Hasegawa, S and Chou, Mei-Yin and Mo, S-K and Fedorov, A-V and others},
  journal={Nature communications},
  volume={8},
  number={1},
  pages={516},
  year={2017},
  publisher={Nature Publishing Group UK London},
  doi={10.1038/s41467-017-00641-1},
  url={https://doi.org/10.1038/s41467-017-00641-1}
}

@article{sky2020theory,
  title={Theory of the thickness dependence of the charge density wave transition in 1 $\mathrm{T-TiTe_2}$},
  author={Sky Zhou, Jianqiang and Bianco, Raffaello and Monacelli, Lorenzo and Errea, Ion and Mauri, Francesco and Calandra, Matteo},
  journal={2D Materials},
  volume={7},
  number={4},
  pages={045032},
  year={2020},
  publisher={IOP Publishing},
  doi={10.1088/2053-1583/abae7a},
  url={https://doi.org/10.1088/2053-1583/abae7a}
}

@article{koley2014preformed,
  title = {{Preformed excitons, orbital selectivity, and charge density wave order in $1T\text{\ensuremath{-}}{\mathrm{TiSe}}_{2}$}},
  author = {Koley, S. and Laad, M. S. and Vidhyadhiraja, N. S. and Taraphder, A.},
  journal = {Phys. Rev. B},
  volume = {90},
  issue = {11},
  pages = {115146},
  numpages = {13},
  year = {2014},
  month = {Sep},
  publisher = {American Physical Society},
  doi = {10.1103/PhysRevB.90.115146},
  url = {https://link.aps.org/doi/10.1103/PhysRevB.90.115146}
}

@article{rossnagel2002charge,
  title={{Charge-density-wave phase transition in $1T\ensuremath{-}{\mathrm{TiSe}}_{2}:$  Excitonic insulator versus band-type Jahn-Teller mechanism}},
  author={Rossnagel, K. and Kipp, L. and Skibowski, M.},
  journal={Phys. Rev. B},
  volume={65},
  issue={23},
  pages={235101},
  numpages={7},
  year={2002},
  month={May},
  publisher={American Physical Society},
  doi={10.1103/PhysRevB.65.235101},
  url={https://link.aps.org/doi/10.1103/PhysRevB.65.235101}
}

@article{weber2011electron,
  title={{Electron-Phonon Coupling and the Soft Phonon Mode in ${\mathrm{TiSe}}_{2}$}},
  author={Weber, F. and Rosenkranz, S. and Castellan, J.-P. and Osborn, R. and Karapetrov, G. and Hott, R. and Heid, R. and Bohnen, K.-P. and Alatas, A.},
  journal={Phys. Rev. Lett.},
  volume={107},
  issue={26},
  pages={266401},
  numpages={5},
  year={2011},
  month={Dec},
  publisher={American Physical Society},
  doi={10.1103/PhysRevLett.107.266401},
  url={https://link.aps.org/doi/10.1103/PhysRevLett.107.266401}
}

@article{hellgren2017critical,
  title={{Critical Role of the Exchange Interaction for the Electronic Structure and Charge-Density-Wave Formation in ${\mathrm{TiSe}}_{2}$}},
  author={Hellgren, Maria and Baima, Jacopo and Bianco, Raffaello and Calandra, Matteo and Mauri, Francesco and Wirtz, Ludger},
  journal={Phys. Rev. Lett.},
  volume={119},
  issue={17},
  pages={176401},
  numpages={6},
  year={2017},
  month={Oct},
  publisher={American Physical Society},
  doi={10.1103/PhysRevLett.119.176401},
  url={https://link.aps.org/doi/10.1103/PhysRevLett.119.176401}
}

@article{novko2022electron,
  title={Electron correlations rule the phonon-driven instability in single-layer {${\mathrm{TiSe}}_{2}$}},
  author={Novko, Dino and Torbatian, Zahra and Lon\ifmmode \check{c}\else \v{c}\fi{}ari\ifmmode \acute{c}\else \'{c}\fi{}, Ivor},
  journal={Phys. Rev. B},
  volume={106},
  issue={24},
  pages={245108},
  numpages={7},
  year={2022},
  month={Dec},
  publisher={American Physical Society},
  doi={10.1103/PhysRevB.106.245108},
  url={https://link.aps.org/doi/10.1103/PhysRevB.106.245108}
}

@article{yang2022coexistence,
  title={Coexistence of the charge density wave state and linearly dispersed energy band in $\mathrm{1T-ZrTe_2}$ monolayer},
  author={Yang, Li-Ning and Xu, Yong-Jie and Li, Qi-Yuan and Meng, Yu-Xin and Zhao, Yi-Fan and Li, Shao-Chun},
  journal={Applied Physics Letters},
  volume={120},
  number={7},
  year={2022},
  publisher={AIP Publishing},
  url={https://doi.org/10.1063/5.0082217}
}

@article{song2023signatures,
  title={Signatures of the exciton gas phase and its condensation in monolayer $\mathrm{1T-ZrTe_2}$},
  author={Song, Yekai and Jia, Chunjing and Xiong, Hongyu and Wang, Binbin and Jiang, Zhicheng and Huang, Kui and Hwang, Jinwoong and Li, Zhuojun and Hwang, Choongyu and Liu, Zhongkai and others},
  journal={Nature communications},
  volume={14},
  number={1},
  pages={1116},
  year={2023},
  publisher={Nature Publishing Group UK London},
  doi={10.1038/s41467-023-36857-7},
  url={https://doi.org/10.1038/s41467-023-36857-7}
}

@article{dey2020structural,
  title={{Structural, electronic, and magnetic properties of vanadium-based Janus dichalcogenide monolayers: A first-principles study}},
  author={Dey, Dibyendu and Botana, Antia S.},
  journal={Phys. Rev. Mater.},
  volume={4},
  issue={7},
  pages={074002},
  numpages={6},
  year={2020},
  month={Jul},
  publisher={American Physical Society},
  doi={10.1103/PhysRevMaterials.4.074002},
  url={https://link.aps.org/doi/10.1103/PhysRevMaterials.4.074002}
}

@article{perdew1996generalized,
  title={{Generalized Gradient Approximation Made Simple}},
  author={Perdew, John P. and Burke, Kieron and Ernzerhof, Matthias},
  journal={Phys. Rev. Lett.},
  volume={77},
  issue={18},
  pages={3865--3868},
  numpages={0},
  year={1996},
  month={Oct},
  publisher={American Physical Society},
  doi={10.1103/PhysRevLett.77.3865},
  url={https://link.aps.org/doi/10.1103/PhysRevLett.77.3865}
}

@article{blochl1994projector,
  title={Projector augmented-wave method},
  author={Bl\"ochl, P. E.},
  journal={Phys. Rev. B},
  volume={50},
  issue={24},
  pages={17953--17979},
  numpages={0},
  year={1994},
  month={Dec},
  publisher={American Physical Society},
  doi={10.1103/PhysRevB.50.17953},
  url={https://link.aps.org/doi/10.1103/PhysRevB.50.17953}
}

@article{kresse1999ultrasoft,
  title={From ultrasoft pseudopotentials to the projector augmented-wave method},
  author={Kresse, G. and Joubert, D.},
  journal={Phys. Rev. B},
  volume={59},
  issue={3},
  pages={1758--1775},
  numpages={0},
  year={1999},
  month={Jan},
  publisher={American Physical Society},
  doi={10.1103/PhysRevB.59.1758},
  url={https://link.aps.org/doi/10.1103/PhysRevB.59.1758}
}

@article{kresse1993ab,
  title={Ab initio molecular dynamics for liquid metals},
  author={Kresse, G. and Hafner, J.},
  journal={Phys. Rev. B},
  volume={47},
  issue={1},
  pages={558--561},
  numpages={0},
  year={1993},
  month={Jan},
  publisher={American Physical Society},
  doi={10.1103/PhysRevB.47.558},
  url={https://link.aps.org/doi/10.1103/PhysRevB.47.558}
}

@article{kresse1996efficiency,
  title={Efficiency of ab-initio total energy calculations for metals and semiconductors using a plane-wave basis set},
  author={Kresse, Georg and Furthm{\"u}ller, J{\"u}rgen},
  journal={Computational materials science},
  volume={6},
  number={1},
  pages={15--50},
  year={1996},
  publisher={Elsevier},
  doi={10.1038/s41467-023-36857-7},
  url={https://doi.org/10.1038/s41467-023-36857-7}
}

@article{giannozzi2009quantum,
  title={{QUANTUM ESPRESSO: a modular and open-source software project for quantum simulations of materials}},
  author={Giannozzi, Paolo and Baroni, Stefano and Bonini, Nicola and Calandra, Matteo and Car, Roberto and Cavazzoni, Carlo and Ceresoli, Davide and Chiarotti, Guido L and Cococcioni, Matteo and Dabo, Ismaila and others},
  journal={Journal of physics: Condensed matter},
  volume={21},
  number={39},
  pages={395502},
  year={2009},
  doi={10.1088/0953-8984/21/39/395502},
  url={https://doi.org/10.1088/0953-8984/21/39/395502}
}

@article{giannozzi2017advanced,
  title={{Advanced capabilities for materials modelling with Quantum ESPRESSO}},
  author={Giannozzi, Paolo and Andreussi, Oliviero and Brumme, Thomas and Bunau, Oana and Buongiorno Nardelli, M and Calandra, Matteo and Car, Roberto and Cavazzoni, Carlo and Ceresoli, Davide and Cococcioni, Matteo and others},
  journal={Journal of physics: Condensed matter},
  volume={29},
  number={46},
  pages={465901},
  year={2017},
  publisher={IOP Publishing},
  doi={10.1088/1361-648X/aa8f79},
  url={https://doi.org/10.1088/1361-648X/aa8f79}
}

@article{giannozzi2020quantum,
  title={{Quantum ESPRESSO toward the exascale}},
  author={Giannozzi, Paolo and Baseggio, Oscar and Bonf{\`a}, Pietro and Brunato, Davide and Car, Roberto and Carnimeo, Ivan and Cavazzoni, Carlo and De Gironcoli, Stefano and Delugas, Pietro and Ferrari Ruffino, Fabrizio and others},
  journal={The Journal of chemical physics},
  volume={152},
  number={15},
  year={2020},
  publisher={AIP Publishing},
  url={https://doi.org/10.1063/5.0005082}
}

@article{garrity2014pseudopotentials,
  title={{Pseudopotentials for high-throughput DFT calculations}},
  author={Garrity, Kevin F and Bennett, Joseph W and Rabe, Karin M and Vanderbilt, David},
  journal={Computational Materials Science},
  volume={81},
  pages={446--452},
  year={2014},
  publisher={Elsevier},
  doi={https://doi.org/10.1016/j.commatsci.2013.08.053},
  url={https://www.sciencedirect.com/science/article/pii/S0927025613005077}
}

@article{mcmillan1968transition,
  title={Transition Temperature of Strong-Coupled Superconductors},
  author={McMillan, W. L.},
  journal={Phys. Rev.},
  volume={167},
  issue={2},
  pages={331--344},
  numpages={0},
  year={1968},
  month={Mar},
  publisher={American Physical Society},
  doi={10.1103/PhysRev.167.331},
  url={https://link.aps.org/doi/10.1103/PhysRev.167.331}
}

@article{allen1975transition,
  title={Transition temperature of strong-coupled superconductors reanalyzed},
  author={Allen, P. B. and Dynes, R. C.},
  journal={Phys. Rev. B},
  volume={12},
  issue={3},
  pages={905--922},
  numpages={0},
  year={1975},
  month={Aug},
  publisher={American Physical Society},
  doi={10.1103/PhysRevB.12.905},
  url={https://link.aps.org/doi/10.1103/PhysRevB.12.905}
}

@article{migdal1958interaction,
  title={Interaction between electrons and lattice vibrations in a normal metal},
  author={Migdal, AB},
  journal={Sov. Phys. JETP},
  volume={7},
  number={6},
  pages={996--1001},
  year={1958},
  url={http://jetp.ras.ru/cgi-bin/dn/e_007_06_0996.pdf}
}

@article{eliashberg1960interactions,
  title={Interactions between electrons and lattice vibrations in a superconductor},
  author={Eliashberg, GM},
  journal={Sov. Phys. JETP},
  volume={11},
  number={3},
  pages={696--702},
  year={1960},
  url={http://jetp.ras.ru/cgi-bin/dn/e_011_03_0696.pdf}
}

@article{ponce2016epw,
  title={{EPW: Electron--phonon coupling, transport and superconducting properties using maximally localized Wannier functions}},
  author={Ponc{\'e}, Samuel and Margine, Elena R and Verdi, Carla and Giustino, Feliciano},
  journal={Computer Physics Communications},
  volume={209},
  pages={116--133},
  year={2016},
  publisher={Elsevier},
  doi={https://doi.org/10.1016/j.cpc.2016.07.028},
  url={https://www.sciencedirect.com/science/article/pii/S0010465516302260}
}

@article{margine2013anisotropic,
  title={Anisotropic Migdal-Eliashberg theory using Wannier functions},
  author={Margine, E. R. and Giustino, F.},
  journal={Phys. Rev. B},
  volume={87},
  issue={2},
  pages={024505},
  numpages={12},
  year={2013},
  month={Jan},
  publisher={American Physical Society},
  doi={10.1103/PhysRevB.87.024505},
  url={https://link.aps.org/doi/10.1103/PhysRevB.87.024505}
}

@article{giustino2007electron,
  title={Electron-phonon interaction using Wannier functions},
  author={Giustino, Feliciano and Cohen, Marvin L. and Louie, Steven G.},
  journal={Phys. Rev. B},
  volume={76},
  issue={16},
  pages={165108},
  numpages={19},
  year={2007},
  month={Oct},
  publisher={American Physical Society},
  doi={10.1103/PhysRevB.76.165108},
  url={https://link.aps.org/doi/10.1103/PhysRevB.76.165108}
}

@UNPUBLISHED{supplementary,
  title={See our supplemental material for computational details and remaining plots including the convergence test, electronic and phonon dispersions under compressive and tensile strain etc.}
 
}

@article{momma2011vesta,
  title={{VESTA 3 for three-dimensional visualization of crystal, volumetric and morphology data}},
  author={Momma, Koichi and Izumi, Fujio},
  journal={Applied Crystallography},
  volume={44},
  number={6},
  pages={1272--1276},
  year={2011},
  publisher={International Union of Crystallography},
  doi={10.1107/S0021889811038970},
  url={https://doi.org/10.1107/S0021889811038970}
}

@article{kawamura2019fermisurfer,
  title={{FermiSurfer: Fermi-surface viewer providing multiple representation schemes}},
  author={Kawamura, Mitsuaki},
  journal={Computer Physics Communications},
  volume={239},
  pages={197--203},
  year={2019},
  publisher={Elsevier},
  doi={https://doi.org/10.1016/j.cpc.2019.01.017},
  url={https://www.sciencedirect.com/science/article/pii/S0010465519300347}
}

@article{Herath2020107080,
title={{PyProcar: A Python library for electronic structure pre/post-processing}},
journal={Computer Physics Communications},
volume={251},
pages={107080},
year={2020},
issn={0010-4655},
doi={https://doi.org/10.1016/j.cpc.2019.107080},
url = {http://www.sciencedirect.com/science/article/pii/S0010465519303935},
author = {Uthpala Herath and Pedram Tavadze and Xu He and Eric Bousquet and Sobhit Singh and Francisco Muñoz and Aldo H. Romero}
}

@article{Lang2024109063,
title={{Expanding PyProcar for new features, maintainability, and reliability}},
journal={Computer Physics Communications},
volume={297},
pages={109063},
year={2024},
issn={0010-4655},
doi={https://doi.org/10.1016/j.cpc.2023.109063},
url={https://www.sciencedirect.com/science/article/pii/S0010465523004083},
author={Logan Lang and Pedram Tavadze and Andres Tellez and Eric Bousquet and He Xu and Francisco Muñoz and Nicolas Vasquez and Uthpala Herath and Aldo H. Romero}
}

@article{peterson2021materials,
  title={Materials discovery through machine learning formation energy},
  author={Peterson, Gordon GC and Brgoch, Jakoah},
  journal={Journal of Physics: Energy},
  volume={3},
  number={2},
  pages={022002},
  year={2021},
  publisher={IOP Publishing},
  doi={10.1088/2515-7655/abe425},
  url={https://doi.org/10.1088/2515-7655/abe425}
}

@article{zhao2025preparation,
  title={{Preparation, Properties, and Applications of 2D Janus Transition Metal Dichalcogenides}},
  author={Zhao, Haoyang and Lam, Jeffrey Chor Keung},
  journal={Crystals},
  volume={15},
  number={6},
  pages={567},
  year={2025},
  publisher={MDPI},
  doi={10.3390/cryst15060567},
  url={https://www.mdpi.com/2073-4352/15/6/567}
}

@article{wang2023decisive,
  title={Decisive role of electron-phonon coupling for phonon and electron instabilities in transition metal dichalcogenides},
  author={Wang, Zishen and Chen, Chuan and Mo, Jinchao and Zhou, Jun and Loh, Kian Ping and Feng, Yuan Ping},
  journal={Phys. Rev. Res.},
  volume={5},
  issue={1},
  pages={013218},
  numpages={11},
  year={2023},
  month={Mar},
  publisher={American Physical Society},
  doi={10.1103/PhysRevResearch.5.013218},
  url={https://link.aps.org/doi/10.1103/PhysRevResearch.5.013218}
}

@article{johannes2008fermi,
  title={Fermi surface nesting and the origin of charge density waves in metals},
  author={Johannes, M. D. and Mazin, I. I.},
  journal={Phys. Rev. B},
  volume={77},
  issue={16},
  pages={165135},
  numpages={8},
  year={2008},
  month={Apr},
  publisher={American Physical Society},
  doi={10.1103/PhysRevB.77.165135},
  url={https://link.aps.org/doi/10.1103/PhysRevB.77.165135}
}

@article{johannes2006fermi,
  title={{Fermi-surface nesting and the origin of the charge-density wave in $\mathrm{Nb}{\mathrm{Se}}_{2}$}},
  author={Johannes, M. D. and Mazin, I. I. and Howells, C. A.},
  journal={Phys. Rev. B},
  volume={73},
  issue={20},
  pages={205102},
  numpages={8},
  year={2006},
  month={May},
  publisher={American Physical Society},
  doi={10.1103/PhysRevB.73.205102},
  url={https://link.aps.org/doi/10.1103/PhysRevB.73.205102}
}

@article{pakdel2025effect,
  title={{Effect of Hubbard U-corrections on the electronic and magnetic properties of 2D materials: a high-throughput study}},
  author={Pakdel, Sahar and Olsen, Thomas and Thygesen, Kristian S},
  journal={npj Computational Materials},
  volume={11},
  number={1},
  pages={18},
  year={2025},
  publisher={Nature Publishing Group UK London},
  doi={10.1038/s41524-024-01503-3},
  url={https://doi.org/10.1038/s41524-024-01503-3}
}

@article{dudarev1998electron,
  title={{Electron-energy-loss spectra and the structural stability of nickel oxide:  An LSDA+U study}},
  author={Dudarev, S. L. and Botton, G. A. and Savrasov, S. Y. and Humphreys, C. J. and Sutton, A. P.},
  journal={Phys. Rev. B},
  volume={57},
  issue={3},
  pages={1505--1509},
  numpages={0},
  year={1998},
  month={Jan},
  publisher={American Physical Society},
  doi={10.1103/PhysRevB.57.1505},
  url={https://link.aps.org/doi/10.1103/PhysRevB.57.1505}
}

@article{imada1998metal,
  title={Metal-insulator transitions},
  author={Imada, Masatoshi and Fujimori, Atsushi and Tokura, Yoshinori},
  journal={Rev. Mod. Phys.},
  volume={70},
  issue={4},
  pages={1039--1263},
  numpages={0},
  year={1998},
  month={Oct},
  publisher={American Physical Society},
  doi={10.1103/RevModPhys.70.1039},
  url={https://link.aps.org/doi/10.1103/RevModPhys.70.1039}
}

@article{feng2022superconductivity,
  title={{Superconductivity induced by Lifshitz transition in pristine $\mathrm{SnS_2}$ under high pressure}},
  author={Feng, Jiajia and Li, Cong and Deng, Wen and Lin, Bencheng and Liu, Wenhui and Susilo, Resta A and Dong, Hongliang and Chen, Zhiqiang and Zhou, Nan and Yi, Xiaolei and others},
  journal={The Journal of Physical Chemistry Letters},
  volume={13},
  number={40},
  pages={9404--9410},
  year={2022},
  publisher={ACS Publications},
  doi={10.1021/acs.jpclett.2c02580},
  url={https://doi.org/10.1021/acs.jpclett.2c02580}
}

@article{wang2024theoretical,
  title={{A theoretical study of Lifshitz transition for $\mathrm{2H-TaS_2}$}},
  author={Wang, Wenxuan and Jiang, Zhenyi and Zhang, Xiaodong and Zheng, Jiming and Du, Hongwei and Zhang, Zhiyong},
  journal={Physical Chemistry Chemical Physics},
  volume={26},
  number={22},
  pages={15868--15876},
  year={2024},
  publisher={Royal Society of Chemistry},
  doi={10.1039/D4CP00977K},
  url={http://dx.doi.org/10.1039/D4CP00977K}
}

\end{document}


\renewcommand{\thefigure}{S\arabic{figure}}
\renewcommand{\thetable}{S\arabic{table}}
\renewcommand{\theequation}{S\arabic{equation}}

\setcounter{figure}{0}
\setcounter{table}{0}
\setcounter{equation}{0}

\vspace*{0.8cm}
\begin{center}
    \large\textbf{Supplementary Material}
\end{center}

\preprint{APS/123-QED}

\title{Observation of intertwined charge density wave order and superconductivity in Janus monolayer}

\author{Subhajit Pramanick\textsuperscript{1,3}}
\email{subhajitbhu@kgpian.iitkgp.ac.in}
\author{Shubham Patel\textsuperscript{2}}
\email{pshubham2805@gmail.com}
\author{Sudip Chakraborty\textsuperscript{3}}
\email{sudipchakraborty@hri.res.in}
\author{A. Taraphder\textsuperscript{1}}
\email{arghya@phy.iitkgp.ac.in}

\affiliation{\textsuperscript{1}Department of Physics, Indian Institute of Technology Kharagpur, Kharagpur 721302, India}
\affiliation{\textsuperscript{2}Department of Physics, University of Bath, Claverton Down, Bath BA2 7AY, United Kingdom}
\affiliation{\textsuperscript{3}Materials Theory for Energy Scavenging Lab, Harish-Chandra Research Institute, A CI of Homi Bhabha National Institute, Chhatnag Road, Jhunsi, Prayagraj 211019, India}

\date{\today}

\maketitle


\section{\label{sec:1}Electronic structures and CDW calculation\protect\\ }

\begin{figure*}[h!]
\includegraphics[width=1.0\linewidth]{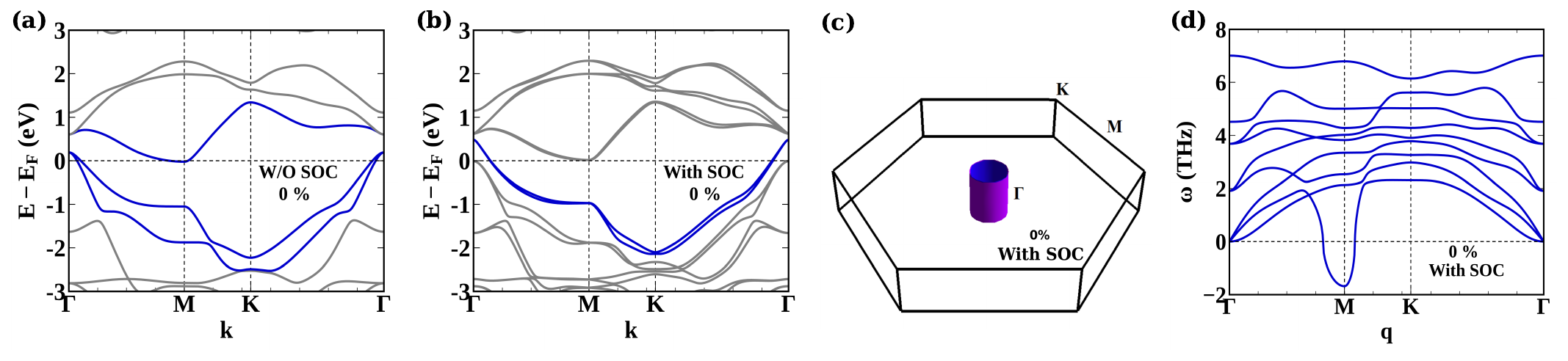}
\caption{\label{fig:1} Electronic band structure of 1 T $\mathrm{ZrSeTe}$ monolayer calculated in QE package \cite{giannozzi2009quantum,giannozzi2017advanced,giannozzi2020quantum} (a) without SOC, (b) with SOC; (c) Fermi surface of 1 T $\mathrm{ZrSeTe}$ monolayer calculated with SOC; (d) Phonon dispersion of 1 T $\mathrm{ZrSeTe}$ monolayer calculated at 0.005 Ry smearing width with SOC.}
\end{figure*}

Electronic band structures corresponding to the ambient conditions computed using the PBE functional \cite{perdew1996generalized} in the Quantum Espresso (QE) package \cite{giannozzi2009quantum,giannozzi2017advanced,giannozzi2020quantum}, both without and with spin-orbit coupling (SOC), are illustrated in Figs. \hyperref[fig:1]{S1(a)} and \hyperref[fig:1]{S1(b)}, respectively. Here, the bands crossing the Fermi level are highlighted in blue, while the remaining bands are shown in gray. The Vienna Ab initio Simulation Package (VASP) \cite{kresse1993ab,kresse1996efficiency} has been used to calculate the electronic dispersion depicted in Fig. 1(c) of the manuscript. It has been employed in the manuscript for the purpose of fatband analysis. Otherwise, all calculations pertaining to electronic band structures (except the electronic bands calculation using DFT+U method and electronic band calculation of CDW distorted phase) and phonon dispersions have been conducted using the QE package \cite{giannozzi2009quantum,giannozzi2017advanced,giannozzi2020quantum}. Nevertheless, all relaxations have been executed within the VASP package \cite{kresse1993ab,kresse1996efficiency}. With the exception of a minor discrepancy in the Fermi energy values, the electronic band structures computed using those two packages are identical.

\begin{figure*}[h!]
\includegraphics[width=1.0\linewidth]{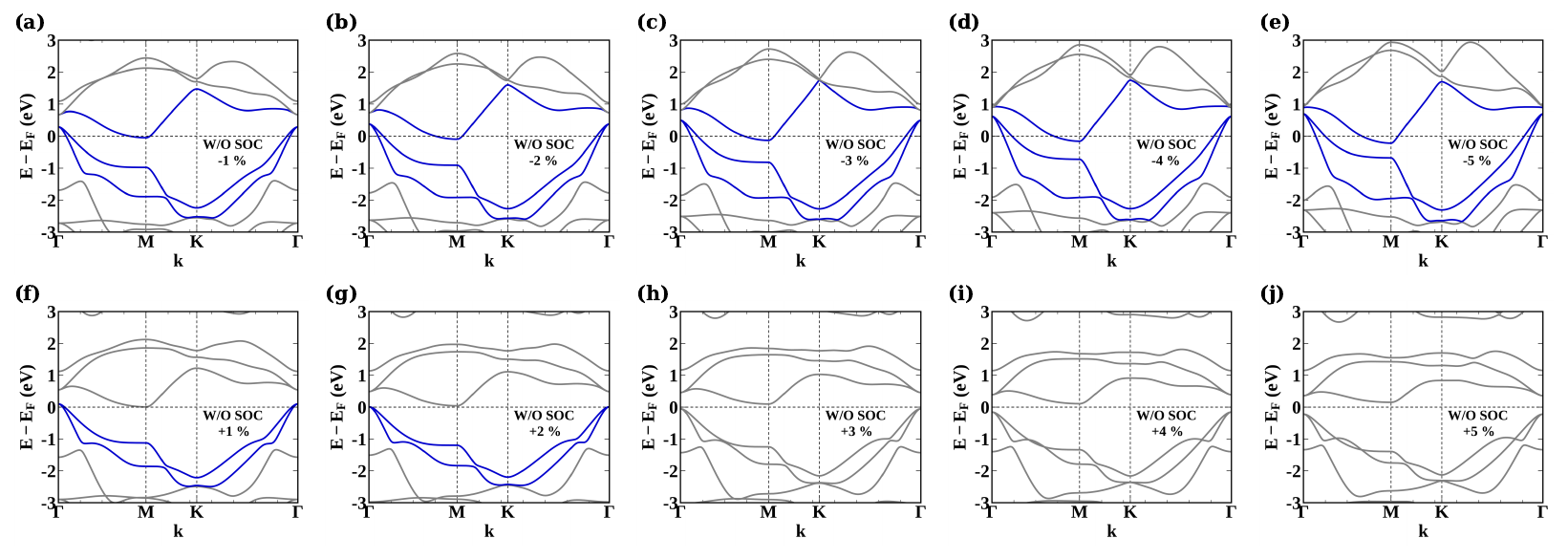}
 \caption{\label{fig:2} Electronic band structure of 1 T $\mathrm{ZrSeTe}$ monolayer for 1 \% to 5 \% (a)-(e) compressive and (f)-(j) tensile strain respectively, calculated without SOC}
\end{figure*}

The presence of SOC results in a reduction of the Fermi energy, as illustrated in Fig. \hyperref[fig:1]{S1(b)}. The impact of SOC on the electronic bands has already been addressed in the manuscript. Figure \hyperref[fig:1]{S1(c)} shows the Fermi surface of the $\mathrm{ZrSeTe}$ monolayer in the presence of SOC. It is clear from this plot that the electron-like pocket at the $\mathrm{M}$ point and the hole-like pocket at the $\mathrm{\Gamma}$ point that are present when SOC is absent now totally vanish. Phonon dispersion calculated with 0.005 Ry smearing width in the presence of SOC has been displayed in the Fig. \hyperref[fig:1]{S1(d)}. The existence of imaginary frequency at the $\mathrm{M}$ point supports the presence of charge density wave (CDW) instability in the presence of SOC. Electronic band structures, Fermi surfaces, and phonon dispersions for different biaxial compressive and tensile strains have been displayed in Fig. \hyperref[fig:2]{S2}-\hyperref[fig:4]{S4} (without SOC case) and Fig. \hyperref[fig:5]{S5}-\hyperref[fig:7]{S7} (with SOC case). The results indicate that compressive strain increases the metallic nature of the 1T $\mathrm{ZrSeTe}$ monolayer. In contrast, tensile strain reduces its metallicity and ultimately leads to a transition from semimetal to semiconductor. The impact of biaxial strain on the electronic properties and CDW instability is already discussed in the manuscript.

\begin{figure*}[h!]
\includegraphics[width=1\linewidth]{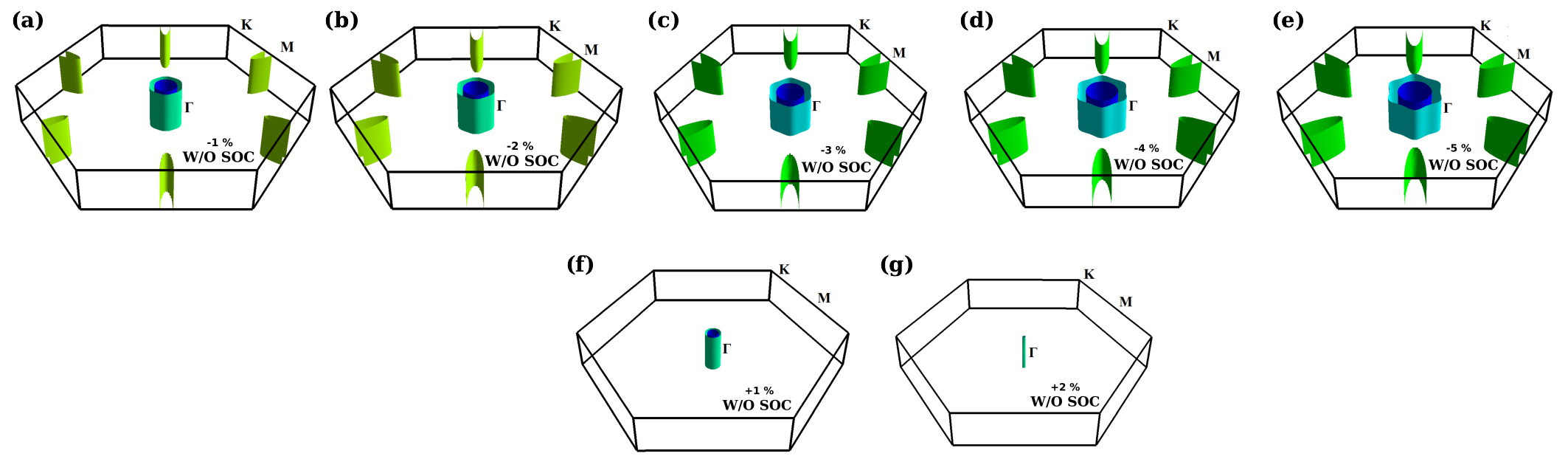}
\caption{\label{fig:3} Fermi surface of 1 T $\mathrm{ZrSeTe}$ monolayer for 1 \% to 5 \% (a)-(e) compressive and 1 \% to 2 \%  (f)-(g) tensile strain respectively, calculated without SOC}
\end{figure*}

\begin{figure*}[h!]
\includegraphics[width=1\linewidth]{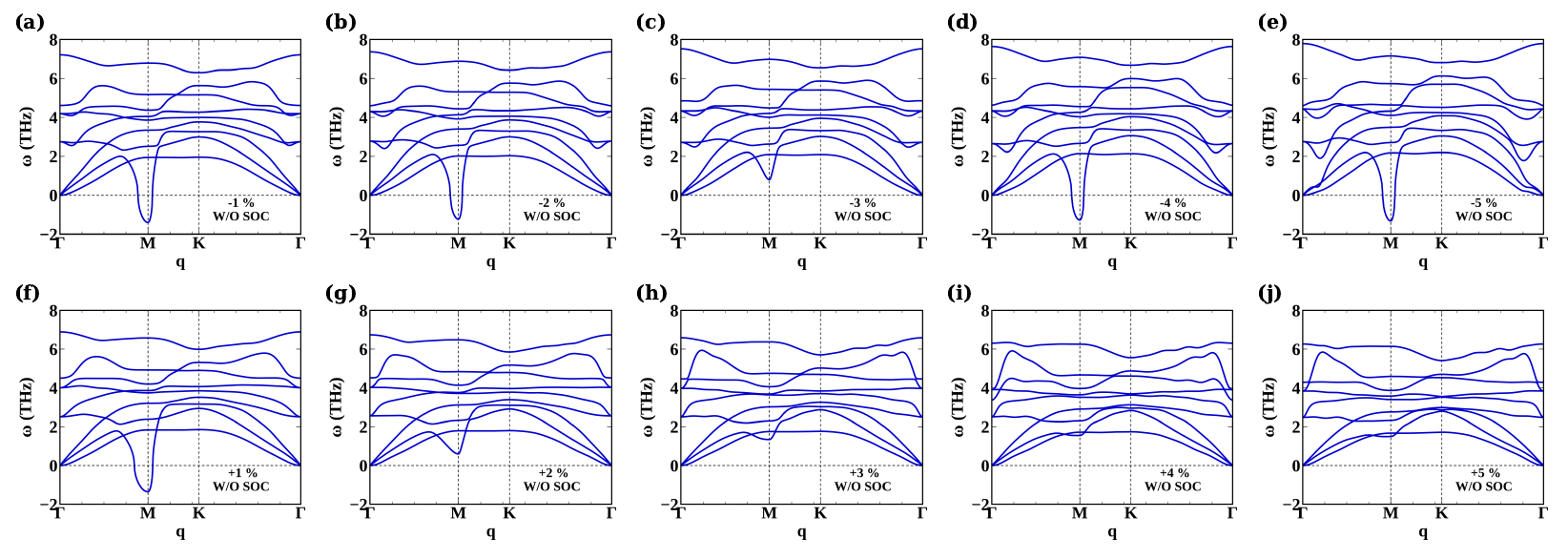}
\caption{\label{fig:4} Phonon band structure of 1 T $\mathrm{ZrSeTe}$ monolayer for 1 \% to 5 \% (a)-(e) compressive and (f)-(j) tensile strain, respectively, calculated at 0.05 Ry smearing width without SOC}
\end{figure*}

\begin{figure*}[h!]
\includegraphics[width=1.0\linewidth]{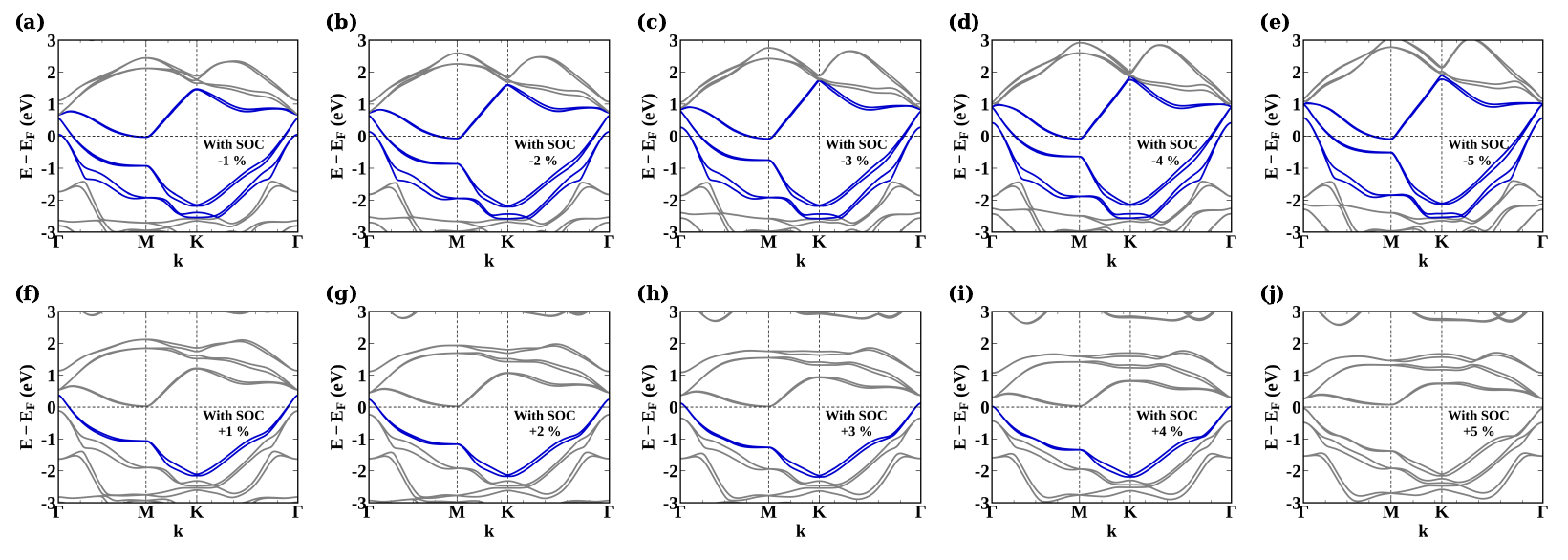}
\caption{\label{fig:5} Electronic band structure of 1 T $\mathrm{ZrSeTe}$ monolayer for 1 \% to 5 \% (a)-(e) compressive and (f)-(j) tensile strain respectively, calculated with SOC}
\end{figure*}

\begin{figure*}[h!]
\includegraphics[width=1.0\linewidth]{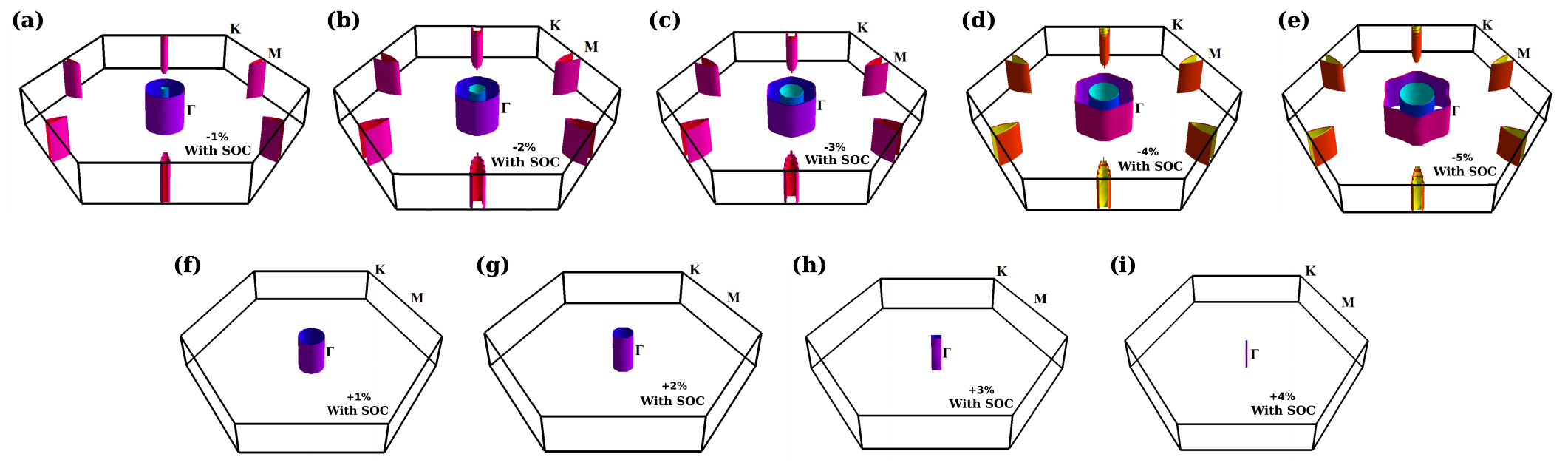}
\caption{\label{fig:6} Fermi surface of 1 T $\mathrm{ZrSeTe}$ monolayer for 1 \% to 5 \% (a)-(e) compressive and 1 \% to 4 \%  (f)-(i) tensile strain respectively, calculated with SOC}
\end{figure*}

\begin{figure*}[h!]
\includegraphics[width=1.0\linewidth]{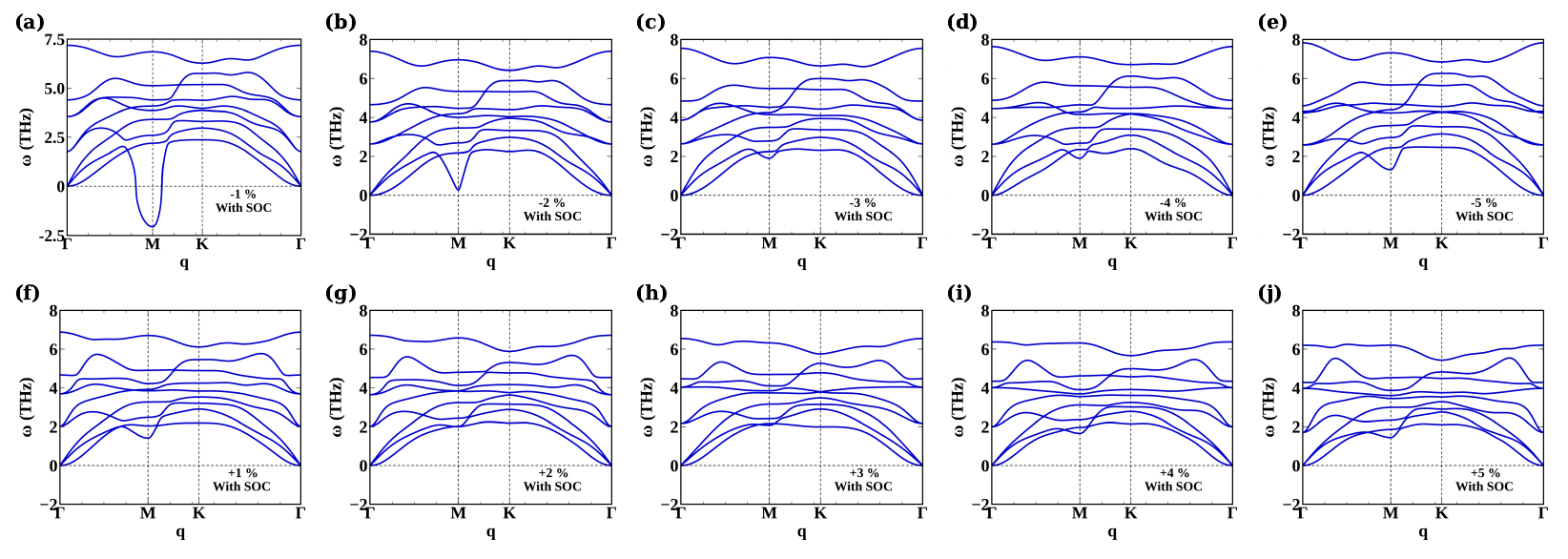}
\caption{\label{fig:7} Phonon band structure of 1 T $\mathrm{ZrSeTe}$ monolayer for 1 \% to 5 \% (a)-(e) compressive and (f)-(j) tensile strain, respectively, calculated at 0.05 Ry smearing width with SOC}
\end{figure*}

\section{\label{sec:2}Superconductivity calculation\protect\\ }

\begin{figure*}[h!]
\includegraphics[width=0.60\linewidth]{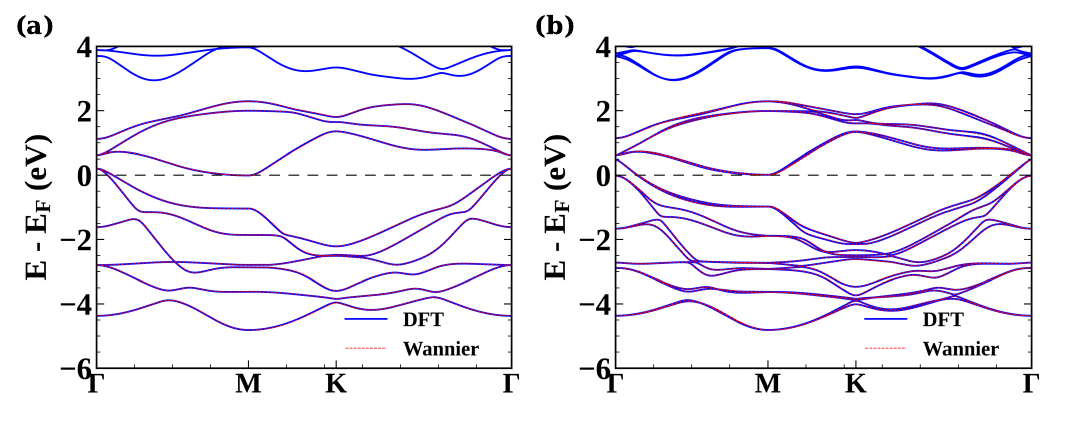}
\caption{\label{fig:8} Comparison between calculated DFT bands and wannier bands (a) in the absence of SOC, (b) in the presence of SOC}
\end{figure*}

We have utilized Migdal-Eliashberg formalism \cite{migdal1958interaction,eliashberg1960interactions} to calculate EPC and superconducting properties using EPW code \cite{ponce2016epw,margine2013anisotropic,giustino2007electron}. For this purpose, wannierization has been done using $\mathit{d_{xz}}$, $\mathit{d_{z^2}}$, $\mathit{d_{x^2-y^2}}$ orbitals of $\mathrm{Zr}$ atom and all \textit{p} orbitals of $\mathrm{Se}$, $\mathrm{Te}$ atoms. Figure \hyperref[fig:8]{S8} displays the comparison between DFT bands and Wannier bands. A good match between DFT bands and Wannier bands ensures the reliability of the Wannier-interpolated electronic structure. Based on this wannierization, other results obtained for EPC and superconductivity have already been discussed in the manuscript.

\section{\label{sec:3}DFT+U calculation\protect\\ }

\begin{figure*}[h!]
\includegraphics[width=0.68\linewidth]{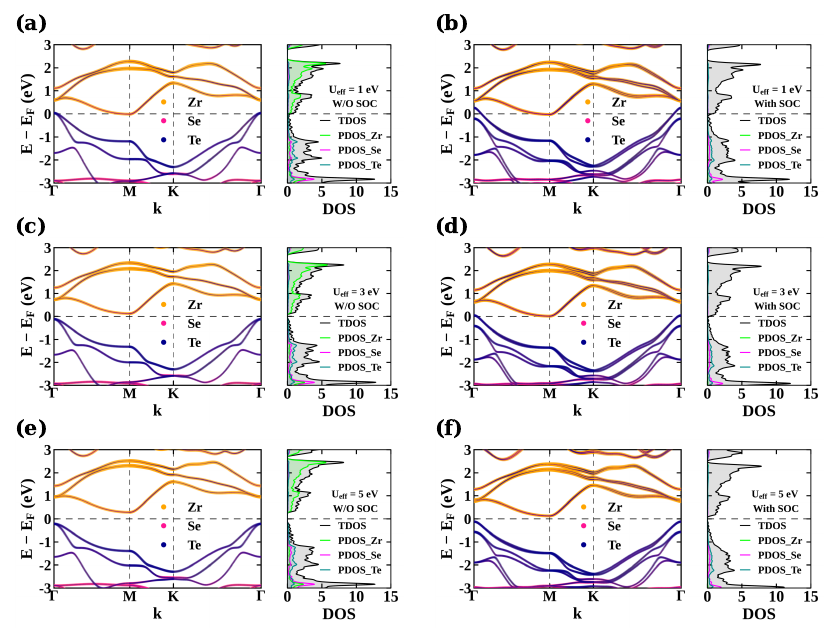}
\caption{\label{fig:9} Electronic band structures and density of states of 1T $\mathrm{ZrSeTe}$ monolayer calculated in the absence of SOC for different values of Hubbard $\mathrm{U_{eff}}$: (a) 1 eV, (c) 3 eV, (e) 5 eV and in the presence of SOC for different values of Hubbard $\mathrm{U_{eff}}$: (b) 1 eV, (d) 3 eV, (f) 5 eV}
\end{figure*}

Electronic structures of 1T $\mathrm{ZrSeTe}$ monolayer for different Hubbard $\mathrm{U_{eff}}$ values, calculated using DFT+U method implemented in VASP package \cite{kresse1993ab,kresse1996efficiency} have been attached in Fig. \hyperref[fig:9]{S9}. These results indicate that the conduction band crossing the Fermi level at the $\mathrm{M}$ point shifts upward, while the valence bands crossing the Fermi level near the $\mathrm{\Gamma}$ point shift downward with increasing $\mathrm{U_{eff}}$. Consequently, the system gradually transitions from a semimetallic to a semiconducting state. The effect of Hubbard $\mathrm{U_{eff}}$ on the electronic properties and CDW distortion has been discussed in detail in the manuscript.

\section{\label{sec:1}Convergence calculation\protect\\ }

\begin{figure*}[h!]
\includegraphics[width=1.0\linewidth]{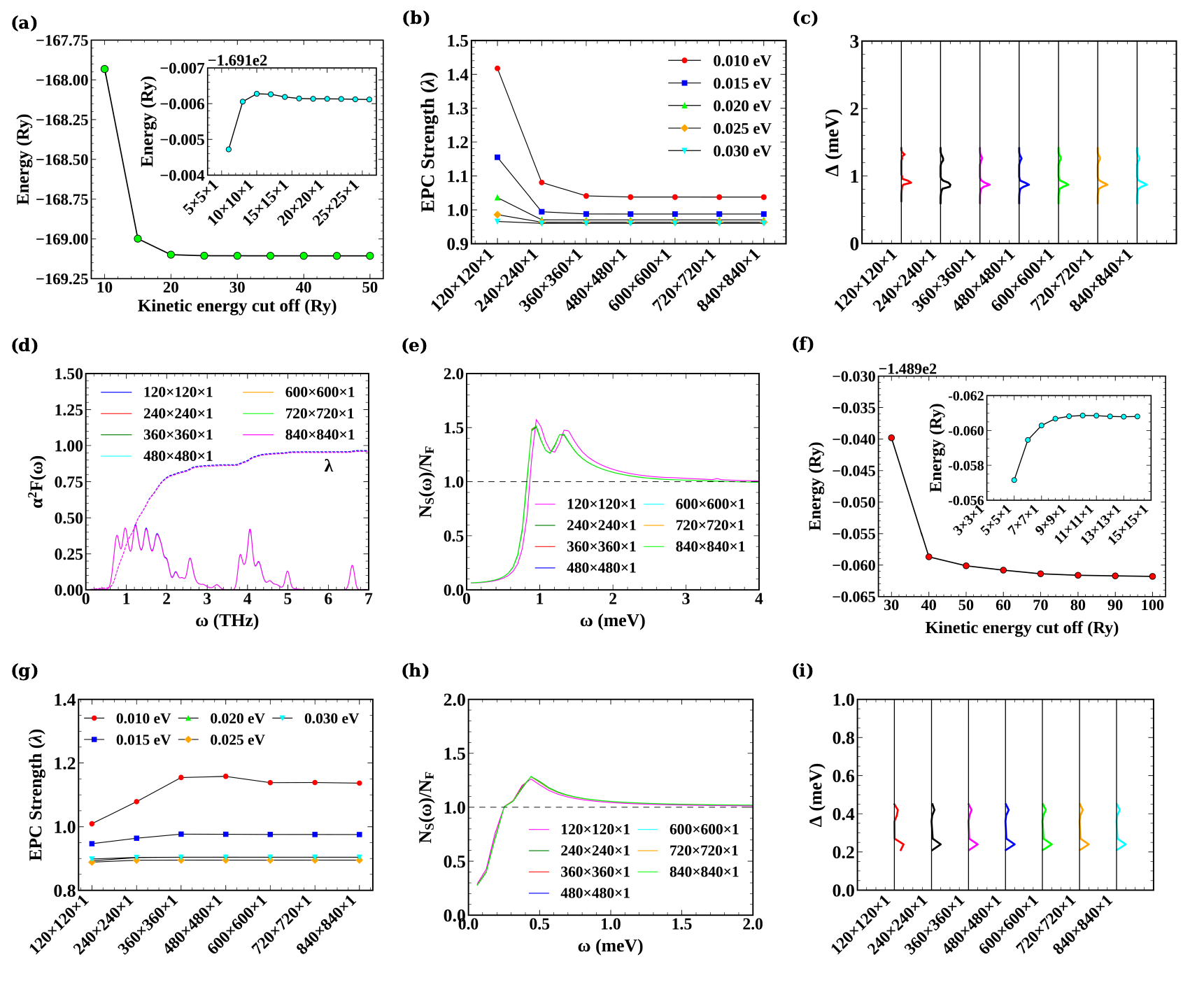}
\caption{\label{fig:10} Convergence test of ground state energy with respect to, Main plot: kinetic energy cut off and Inset plot: k-points grid (a) in absence of SOC, (f) in presence of SOC; Convergence test of EPC strength with respect to fine k-grid for different values of smearing width (b) in absence of SOC, (g) in presence of SOC; Gap convergence with respect to different fine k-grids calculated for 0.03 eV smearing width (c) in absence of SOC, (i) in presence of SOC; (d) Convergence test of Eliashberg spectral function with respect to different fine k-grids calculated at 0.03 eV electronic smearing width; Convergence test of superconducting density of states with respect to different fine k-grids calculated at 0.03 eV smearing width (e) in absence of SOC, (h) in presence of SOC. For (c) and (e), a 4 K temperature has been used, and for (h) and (i), a 1 K temperature has been used.}
\end{figure*}

Figure \hyperref[fig:10]{S10(a)} shows the convergence behavior of the ground state energy (without SOC) with respect to the plane wave kinetic energy cutoff and \textit{k}-point mesh in QE package \cite{giannozzi2009quantum,giannozzi2017advanced,giannozzi2020quantum}, indicating that a cutoff of 30 Ry and a 24$\times$24$\times$1 \textit{k}-grid are sufficient for convergence. The convergence of EPC strength with respect to fine \textit{k}-point grids and electronic smearing widths, calculated using EPW \cite{ponce2016epw,margine2013anisotropic,giustino2007electron} (without SOC), is presented in Fig. \hyperref[fig:10]{S10(b)}. The results indicate that an electronic smearing of 0.03 eV together with a 600$\times$600$\times$1 fine \textit{k}-grid yields stable EPC values. Using this smearing width, the superconducting gap, Eliashberg spectral function, and superconducting density of states have been computed (without SOC) for several fine \textit{k}-grids shown in Figs. \hyperref[fig:10]{S10(c–e)} respectively, confirming that the 600$\times$600$\times$1 fine \textit{k}-grid provides converged results.

Similarly, Fig. \hyperref[fig:10]{S10(f)} presents the convergence of the ground state energy (with SOC) with respect to the kinetic energy cutoff and \textit{k}-point mesh, showing that 70 Ry and a 12$\times$12$\times$1 \textit{k}-grid are sufficient for convergence. Figure \hyperref[fig:10]{S10(g)} shows the convergence of EPC strength with respect to fine \textit{k}-grids and electronic smearing widths in EPW \cite{ponce2016epw,margine2013anisotropic,giustino2007electron} (with SOC), indicating that a smearing of 0.03 eV and a 600$\times$600$\times$1 fine \textit{k}-grid provide reliable results. Using this smearing width value, the superconducting density of states and superconducting gap have been calculated (with SOC) for different fine \textit{k}-grids in Figs. \hyperref[fig:10]{S10(h-i)} respectively, which verify that 600$\times$600$\times$1 is sufficient for converged results.

\bibliography{SM}